\documentclass[12pt, draftclsnofoot, onecolumn]{IEEEtran}
\usepackage{amsthm,amsmath,amssymb,mathtools,bm,etoolbox,float}
\usepackage{graphicx,cite,cuted}
\usepackage{bigints}
\usepackage{caption}

\newcommand{\erf}{\operatorname{erf}}

\usepackage{stackengine}

\usepackage{ellipsis}
\usepackage{tkz-euclide}
\usepackage{tikz}
\usepackage{tikz-3dplot}
\usepackage[utf8]{inputenc}
\usepackage{pgfplots} 
\usepackage{pgfgantt}
\usepackage{pdflscape}
\pgfplotsset{compat=newest} 
\pgfplotsset{plot coordinates/math parser=false}
\pgfplotsset{every  tick/.style={black,},ylabel style={font=\tiny},xlabel style={font=\tiny},tick label style={font=\tiny},legend style= {font=\scriptsize},
minor x tick num=1,minor y tick num=1,xminorticks=true,yminorticks=true,}
  \newlength\fheight
\newlength\fwidth
\begin{document}

%\title{MmWaves Vehicular Cooperative Diversity Relay Fast Fading Channels with Interference and Blockage Constraints}
%\title{MmWaves Vehicular Cooperative Diversity Relay Time-Selective Fading with Interference and Blockage Constraints}
\title{MmWaves Cellular V2X for Cooperative Diversity Relay Fast Fading Channels}

\author{Elyes Balti,~\IEEEmembership{Member,~IEEE}, and Brian K. Johnson,~\IEEEmembership{Senior~Member,~IEEE}}

%\thanks{Y. Fang is with the Department of Electrical and Computer Engineering, University of Florida, Gainesville, FL, 32611, USA e-mail: (see http://winet.ece.ufl.edu/tvt/ for further information regarding IEEE TVT.)}% <-this % stops a space
%\thanks{Manuscript received XXX, XX, 2019; revised XXX, XX, 2019.}}

%\markboth{IEEE Transactions on Vehicular Technology,~Vol.~XX, No.~XX, XXX~2019}
%{}
%{Shell \MakeLowercase{\textit{et al.}}: Bare Demo of IEEEtran.cls for Journals}

\maketitle

\begin{abstract}
In this work, we present a framework analysis of millimeter waves (mmWaves) vehicular communications systems. Communications between vehicles take place through a cooperative relay which acts as an intermediary base station (BS). The relay is equipped with multiple transmit and receive antennas and it employs decode-and-forward (DF) to process the signal. Also, the relay applies maximal ratio combining (MRC), and maximal ratio transmission (MRT), respectively, to receive and forward the signal.
As the vehicles' speeds are relatively high, the channel experiences a fast fading and this time variation is modeled following the Jakes' autocorrelation model. We also assume narrowband fading channel. Closed-form expressions of the reliability metrics such as the outage probability, the probability of error and the channel capacity are derived. Capitalizing on these performances, we derive the low and high power regimes for the capacity, and the high signal-to-interference-plus-noise-ratio (SINR) asymptotes for the outage and error probability to get full insights into the system gains such as the diversity and coding gains.
\end{abstract}

\begin{IEEEkeywords}
Millimeter waves, cellular, vehicular communications, relay diversity, fast fading.
\end{IEEEkeywords}

\IEEEpeerreviewmaketitle

\section{Introduction}

%\IEEEPARstart{}{}
The increase demand for bandwidth has been growing over the last few decades, largely due to the increased number of subscribers and the number of communication devices. Due to these factors, the users suffer from the spectrum scarcity as most of the available spectrum is totally assigned to the licensed users. In addition, spectrum sharing-sensing systems which consist of primary and secondary users also reach its bottleneck. Due to the massive number of users, the secondary users become unable to take advantages of the spectrum holes left by the primary users. Also, the primary users are not efficiently communicating since the licensed microwave spectrum becomes very limited. Furthermore, advanced Long Term Evolution (LTE) communications systems are in desperate needs for high-speed wireless data rate in order to share big data, high definition (HD) videos, e.g. advanced driving assistance system (ADAS) for vehicular communications \cite{v1,v2,v3}. Self-driving vehicles require a big amount of shared data with the other surrounding vehicles to help the vehicles exploring their environments and reduce the rates of accidents.

To address these challenges, millimeter waves (mmWaves) technology has been emerged as a promising solution to overcome these shortcoming. In fact, mmWaves provides not only a large band of available spectrum, but also a high data rate for exchanging the data. Furthermore, the capacity of the wireless cellular network massively increases and become able to support a large number of subscribers compared to the microwave cellular systems. Hence, mmWaves technology is the best way to densify the cellular network and support the technologies that require high data rates \cite{tractable}.
\subsection{Motivation}
mmWaves communications which refer to the wide spectrum between 28 and 300 GHz become a promising solution to provide high Gbps of bandwidth to emerging 5G network \cite{1,5}. 
Such interesting technology has recently attracted enormous attentions both in academia and industry \cite{2}. HD video applications leverage from mmWaves frequencies to transmit ultra HD video to HDTV wirelessly \cite{3}. In addition, mmWaves can also replace fiber optic transmission lines connecting the core network to the backhaul when the cable installation is restricted or expensive \cite{8}. Since mmWaves spectrum becomes available for use, industries invested and produced mmWaves based commercial products such as IEEE 802.11ad wireless gigabit alliance (WiGig) technology. In fact, this product is dedicated to support wireless display interfaces requiring high resolution and gigabit rate \cite{2,3,4}. Furthermore, mmWaves frequencies support radar applications for motion sensors, automatic doors, and collision avoidance systems \cite{3,6}. Moreover, mmWaves communications have low interference and high security against hacking and intrusion. In fact, mmWaves link has narrow beam and short range where the area experiences less interference from neighbor radios. Also, mmWaves system is immune since the propagation is restricted to a limited area which enhances the security \cite{3,v4,v5}. In the same context, mmWaves technology has tremendous applications in vehicular communications area. Specifically, self-driving vehicles require a great amount of information from nearby vehicles and road side units (RSU) to enhance the vehicle awareness and avoid the potential accidents. Certainly such application requires a high bandwidth and data rate to support the big data exchanged between the vehicles and RSUs \cite{6,7,maalej}.

Although mmWaves technology has tremendous advantages, it suffers from many drawbacks. Due to the small wavelength, mmWaves has a short propagation range around 20 meters for low power applications \cite{9}. Supporting longer ranges requires large antenna gains and transmit power. Also, since mmWaves frequencies are sensitive to blockage, it requires LOS communication as regular objects such as trees, human being, and building completely block the signal \cite{5}. Consequently, the range is reduced and the received signal is weak. Furthermore, due to the high frequencies, mmWaves is substantially degraded by the path loss as it suffers from the atmospheric attenuations such as the moisture, fog, rain attenuations, and the oxygen absorption \cite{10}. To compensate for the path loss, mmWaves systems use high directional array antennas \cite{antenna}. Moreover, due to the large bandwidth, the received mmWaves signal experiences high noise degradations at the receiver which substantially limits the received signal-to-noise ratio (SNR). 

To enhance the quality of the received signal, relay based cooperative communication technique has been proposed not only to improve the capacity but also to improve the coverage and assist the received signal power \cite{11,r1,r2,eT}. Related works have proposed various relaying techniques to process the signal such as amplify-and-forward (AF), and decode-and-forward (DF) \cite{c1,c2,c3,j1,j4,12,13}. The AF consists of amplifying the received signal with a proper gain factor and then forwards it, while DF decodes and denoises the signal and then forwards it. Since mmWaves are susceptible to the high noise power, the AF is not relevant because in this case the relay also amplifies the noise which is incorporated in the received signal. Besides, research attempts have introduced the full-duplex relaying as it has the potential to double the spectral efficiency. Due to the self-interference which substantially degrades the performance, related works proposed beamforming designs to cancel the self-interference and improve the achievable rate \cite{fd1,fd2}. Furthermore, relaying as well as the intelligent reflecting surfaces (IRS) have been introduced in the context of physical layer security to protect the legitimate receiver from the eavesdropping attacks by transmitting friendly jammers and/or artificial noise in order to maximize the secrecy capacity \cite{pls1,pls2}.
\subsection{Related works}
Tremendous work dealing with cooperative relaying communication have been published. P. Trinh \emph{et.~al} \cite{e1} have proposed a mmWaves cellular system with free space optical (FSO) backhauling. The mmWaves channel follows the Rician model to characterize the LOS component. The outcomes of this work are the outage probability, the error rate, and the channel capacity. Furthermore, M.~Makki \emph{et.~al} \cite{e2} developed a framework analysis of hybrid mmWaves and FSO DF relaying system for multiple hop and mesh topologies. They concluded that Hybrid Automatic Repeat reQuest (HARQ) improves the power efficiency and compensate for the impact of the hardware impairments. For multiple hop topology, the outage is not sensitive to the large number of transmit antennas while the capacity substantially depdends on the number of antennas. In addition, M. Ozpolat \emph{et.~al} \cite{e3} provided a grid-based region coverage analysis of urban mmWaves vehicular ad hoc network (VANET) \cite{maalej,neji1,neji2}. The results demonstrated that mmWaves enables VANET to support fully connected traffic and to satisfy the needs for a dense traffic scenario. Moreover, Z. Sheng \emph{et.~al} \cite{e4} considered the performance analysis of a 5G vehicular network with reinforcement learning-based dedicated short range communications (DSRC) for vehicle to vehicle and mmWaves for vehicle to infrastructure communications. The main idea of this work is to integrate both DSRC and mmWaves in vehicular network and to manage the connectivity components and a situational-aware central unit to optimize the handoff decisions. 
\subsection{Contribution}
In this context, we propose a dual-hop vehicular system wherein the relay employs DF relaying mode. The source and the destination are both a transmit (TX) and receive (RX) vehicles while the relay behaves as an intermediary base station (BS). We adopt the Maximal ratio combining (MRC) at the receiver side of the relay, while the maximal ratio transmission (MRT) is considered at the transmitter side of the relay. Due to the high mobility of the vehicles, the channels are considered time-varying (fast fading) at the first and the second hops. Consequently, the TX and RX are not able to track the channel variations and get a perfect estimate of the channels' coefficients. We assume the Jake's autocorrelation to model the relation between the updated and outdated channel state information (CSI). In addition, we consider the interferences, respectively, from the nearby vehicles for the uplink scenario (first hop), and from the nearby BSs for the downlink scenario (second hop). Furthermore, we assume that the propagating signals experience the blockage since randomly distributed objects such as the buildings, birds, and human bodies can exist between the TX, RX, and the BSs. We also consider the free space path loss model encompassing the rain attenuation, and the oxygen absorption. Nakagami-$m$ model is proposed for modeling the channel fading between the Tx, the relaying BSs, and the Rx. The analysis of the system performance follows these steps:
\begin{itemize}
    \item Provide the analysis of the system and channels' models for a single hop, since the two hops are symmetric.
    \item Provide the statistical distributions of the signal-to-interference-plus-noise (SINR) for a single hop.
    \item Derive the system performance metrics for a single hop.
    \item Capitalizing on these metrics, we derive the high SINR characteristics in order to obtain the system gains such as the diversity and the coding gain for a single hop.
    \item Generalize the analysis of the performance metrics for the overall relaying system.
\end{itemize}
\subsection{Structure}
This paper is organized as follows: the system and channel model are analyzed in Section II. The reliability metrics analysis for a single hop are detailed in Section III, while the performance analysis of the overall relaying system is reported in Section IV. Section V presents the discussion of the numerical results and the concluding remarks are reported in Section VI. 
\section{System and Channel Model}
\subsection{System Description}
\begin{figure}[H]
\centering
\includegraphics[scale=1]{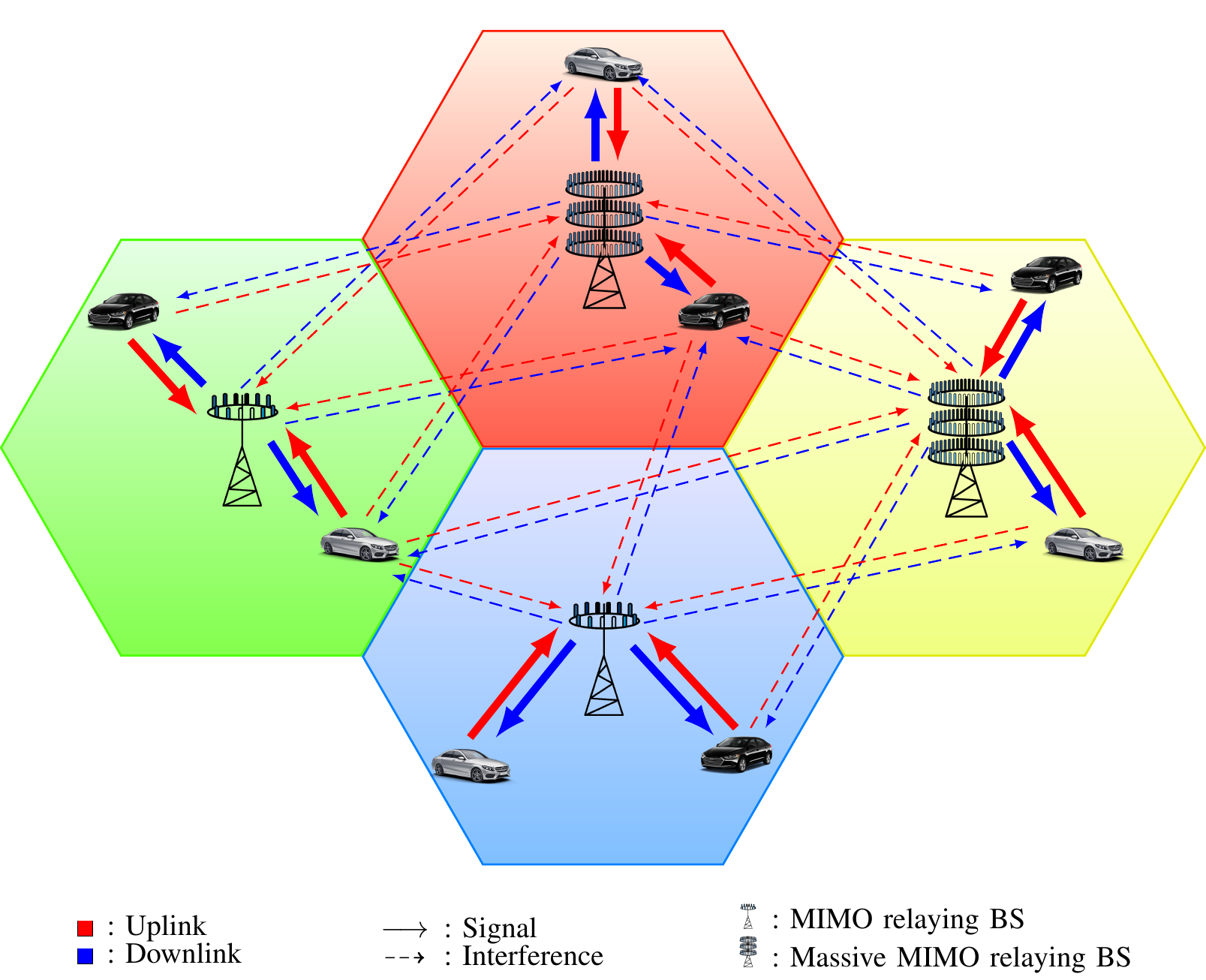}
    \caption{mmWaves uplink and downlink 5G cellular cooperative network architecture. The transmit and receive vehicles are communicating through the BSs.}
    \label{fig1}
\end{figure}

Fig.~\ref{fig1} provides an illustration of the relay network. The proposed system consists of two vehicles communicating through an intermediary relaying BS. Each vehicle is equipped with a single antenna while the relay is equipped with multiple TX and RX antennas, and employs DF protocol to process the received signal. We assume that the RX antennas elements are close enough and the distance between the relay and TX vehicle is long so that the angle of arrival and the time delay are the same for all the signals. The same assumption holds for the second hop, where the transmitted signals at the relay have the same angle of departure $\phi_2$ and time delay to reach the receive vehicle. To estimate the channel state information (CSI), the relay sends periodic feedback to the vehicles. Since the feedback is slowly propagating and the channels are time-varying, the transmitted CSI will be the outdated version. The time correlation between the outdated and updated CSIs can be modeled using the Jakes' autocorrelation model as follows \cite[Eq.~(8)]{14}
\begin{equation}{\label{eq1}}
\rho_i = [J_0(2\pi f_{d,i} \tau_{d,i})]^2
\end{equation}
where $J_0(\cdot)$ is the first kind of Bessel function with order zero, the index $i = 1,~2$ stands for the first and second hop. $f_{d,i} = \frac{f_cv_i\cos(\phi_i)}{c}$ is the Doppler frequency, $\phi$ is the angle between vehicle displacement and the relay, $f_c$ is the carrier frequency, $v_i$ is the speed of the vehicle, $c$ is the speed of light, and $\tau_{d,i}$ is the propagation delay.
%\begin{figure}[H]
%\centering
%\setlength\fheight{6cm}
%\setlength\fwidth{12cm}
%\input{cellular.tikz}
%    \caption{mmWaves uplink and downlink 5G cellular cooperative network architecture. The transmit and receive vehicles are communicating through the BSs.}
%    \label{fig1}
%\end{figure}

\begin{figure}[H]
\centering
\setlength\fheight{5cm}
\setlength\fwidth{7cm}
\input{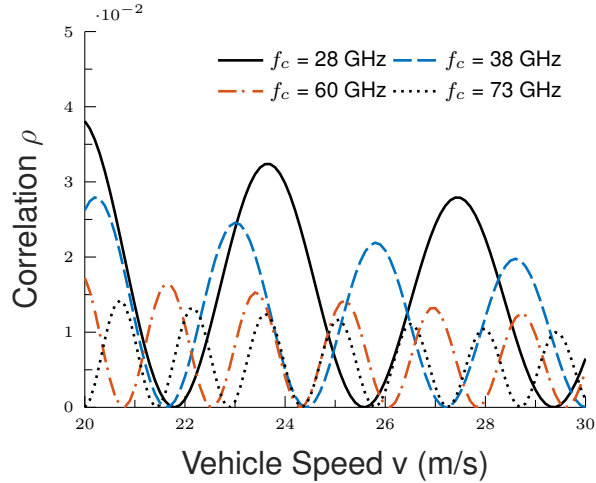}
    \caption{Correlation dependence on the vehicle speed for different values of the carrier frequency. The symbol period $T_s$ = 1 ms, $\tau_d$ = 2$T_s$, and $\phi = \frac{\pi}{4}$.}
     \label{fig2}
\end{figure}

Fig.~\ref{fig2} illustrates the time correlation with respect to the frequency and the vehicle speed. The maximum coherence time $T_c$ corresponding to 28 GHz and a speed of 10 m/s, is 757.61 $\mu$s, while the minimum $T_c$ corresponding to 73 GHz and a speed of 30 m/s is 193.73 $\mu$s. We observe that the correlation peaks decrease with the speed and the carrier frequency. As the vehicle speed increases mainly in highways and rural areas, the correlation decreases and inversely when the vehicle speed is lower, specifically in urban and sub-urban areas, the correlation improves. In fact, for higher speeds, the channel is rapidly varying and the estimation becomes worse. However, for lower speed, the TX and RX can relatively track the channel variations, and hence the estimation gets much better.

The relation between the outdated $\hat{\gamma_{i}}$ and updated $\gamma_{i}$ instantaneous SNRs of the hop $i$ is expressed as follows \cite[Eq.~(7)]{14}
\begin{equation}{\label{eq2}}
\hat{\gamma}_{i} = \sqrt{\rho_i}\gamma_{i} + \sqrt{1 - \rho_i}\omega_i
\end{equation}
where $\omega_i$ is complex Gaussian distributed with zero-mean and having the same variance as $\gamma_i$.
\subsection{MmWaves channel model of the hop $i$}
Since the channels of the first and second hops are symmetric, and the relay transmit and receive strategies are also equivalent (MRC and MRT), we will focus the analysis on a single hop.
\begin{equation}\label{eq3}
y_{i} = \sqrt{\frac{\Omega_{i}P_i}{N_i}}\sum_{n=1}^{N_i}h_{n,i}s + \sum_{k=1}^{M_{r,i}}f_{k,i}d_{k,i} + n_{i},
\end{equation}
where $\Omega_i$ is the average power gain, $P_i$ is the total transmitted power equally splitted among the $N_i$ channels, $s$ is the modulated symbol, $h_{n,i}$ is the $n$-th channel, $d_{k,i}$ is the modulated symbol of the $k$-th interferer with an average power $\mathbb{E}[|d_{k,i}|^2] = P_{k,i}$, $f_{k,i}$ is the fading coefficient of the $k$-th interferer, $M_{r,i}$ is the number of the interferers, and the noise $n_i$ is a zero-mean additive white Gaussian noise (AWGN). The noise variance is given by
\begin{equation}\label{eq4}
\sigma_i^2[\text{dBm}] = B + N_0 + N_f.
\end{equation}  
The path gain of the link $i$ is given by \cite{10}
\begin{equation}\label{eq5}
%\begin{split}
%\PL_i[\text{dB}] =& 20\log_{10}\left(\frac{4\pi f_c}{c}\right) + 10\sigma\log_{10}(L_i) - G_{t,i} - G_{r,i},
%\end{split}
\Omega_i = G_{t,i} + G_{r,i} -20\log_{10}\left(\frac{4\pi L_i f_c}{c} \right) - (\alpha_{\text{ox}} + \alpha_{\text{rain}})L_i
\end{equation} 
where $\alpha_{\text{ox}}$ and $\alpha_{\text{rain}}$ are the oxygen absorption, and the rain attenuation coefficients, respectively.

TABLE (\ref{params}) summarizes the values of the system parameters used in the simulation. 
\begin{table}[H]
\renewcommand{\arraystretch}{1}
\caption{System Parameters}\label{params}
\label{table_example}
\centering
\begin{tabular}{|c|c|c|}
\hline
\bfseries Parameter & \bfseries Symbol  &  \bfseries Value \\
\hline
Transmit antenna element gain & $G_{t,i}$   & 44 dB\\
\hline
Receive antenna element gain & $G_{r,i}$   &  44 dB\\
\hline
Noise spectral density & $N_0$   & -142 dBm/Hz\\
\hline
Noise figure & $N_f$   & 0 dB\\
\hline
Bandwidth & $B$   & 700 MHz\\
\hline
Speed of light & $c$   & 3 $10^8$ m/s\\
\hline
Link distance & $L_i$   & 50 m\\
\hline
\end{tabular}
\end{table}
\section{Performance Analysis of Hop $i$}
\subsection{SINR statistical analysis}
Given that the single input single output (SISO) channel fading is Nakagami-m distributed, the relative SNR follows the Gamma distribution. Hence the overall SNR, which is the sum of $N_i$ Gamma identically distributed SNR per branch with scale $m_i$, is also a Gamma random variable with scale $N_im_i$.

The joint probability density function (PDF) of the outdated and updated SNRs is given by \cite[Eq.~(14)]{14}
\begin{equation}\label{eq6}
\begin{split}
f_{\hat{\gamma_i},\gamma_i}(x,y) = &\left( \frac{N_im_i}{\overline{\gamma}_i}\right)^{N_im_i+1}\frac{\left(\frac{xy}{\rho_i}\right)^{\frac{N_im_i-1}{2}}}{(1-\rho_i)\Gamma(N_im_i)}e^{-\left( \frac{x+y}{1-\rho_i}\right)\frac{N_im_i}{\overline{\gamma}_i}}I_{N_im_i-1}\left(\frac{2N_im_i\sqrt{\rho_ixy}}{(1-\rho_i)\overline{\gamma}_i}\right)
\end{split}
\end{equation} 
where $I_{\nu}(\cdot)$ is the $\nu$-th order modified Bessel function of the first kind and $\overline{\gamma}_i$ is the average SNR. After some mathematical manipulations, the PDF of the outdated CSI is obtained by
\begin{equation}\label{eq7}
f_{\hat{\gamma_i}}(x) = \left(\frac{N_im_i}{\overline{\gamma}_i}\right)^{N_im_i}\frac{x^{N_im_i}-1}{\Gamma(N_im_i)} e^{-\frac{N_im_ix}{(1-\rho_i)\overline{\gamma}_i}}.
\end{equation}
Integrating (\ref{eq7}) and using \cite[Eq.~(3.351.1)]{15}, the cumulative distribution function (CDF) is given by 
\begin{equation}\label{eq8}
F_{\hat{\gamma_i}}(x) = \frac{(1-\rho_i)^{N_im_i}}{\Gamma(N_im_i)} \gamma\left(N_im_i,~\frac{N_im_ix}{(1-\rho_i)\overline{\gamma}_i} \right)
\end{equation}
where $\gamma(\cdot,\cdot)$ is the incomplete lower gamma function. Since $N_im_i$ is an integer, the CDF (\ref{eq8}) can be reformulated as follows
\begin{equation}\label{eq9}
\begin{split}
F_{\hat{\gamma_i}}(x) = (1-\rho_i)^{N_im_i} \left(1 - e^{-\frac{N_im_ix}{(1-\rho_i)\overline{\gamma}_i}}\sum_{n=0}^{N_im_i-1}\frac{x^n}{n!} \left(\frac{N_im_i}{(1-\rho_i)\overline{\gamma}_i}\right)^{n} \right).
\end{split}
\end{equation} 
We assume that each interferer fading follows Nakagami-m with parameter $m_{r,i}$ and we neglect the impact of the time correlation on each channel. Repeating the same derivation steps, the PDF of the aggregate $M_{r_i}$ interferers' SNR $\gamma_{r,i}$ is given by
\begin{equation}\label{eq10}
f_{\gamma_{r,i}}(x) = \left(\frac{M_{r,i}m_{r,i}}{\overline{\gamma}_{r,i}}\right)^{M_{r,i}m_{r,i}} \frac{x^{M_{r,i}m_{r,i}-1}}{\Gamma(M_{r,i}m_{r,i})}e^{-\frac{M_{r,i}m_{r,i}x}{\overline{\gamma}_{r,i}}}
\end{equation}
where $\overline{\gamma}_{r,i}$ is the average SNR of the interferers. 

The effective SINR is expressed by
\begin{equation}\label{eq11}
\gamma_{\text{eff},i} = \frac{\hat{\gamma}_{i}}{\gamma_{r,i} + 1}.
\end{equation} 
Using \cite[Eq.~(3.351.3)]{15}, and after some mathematical manipulations, the CDF of the effective SINR is given by
\begin{equation}\label{eq12}
\begin{split}
&F_{\gamma_{\text{eff},i}}(x) = (1-\rho_i)^{N_im_i} \left[1 - \left(\frac{M_{r,i}m_{r,i}}{\overline{\gamma}_{r,i}}\right)^{m_{r,i}} \frac{1}{\Gamma(M_{r,i}m_{r,i})} \right.\\&~\times\left.
\sum_{n=0}^{N_im_i-1}\sum_{p=0}^{n} \frac{{n \choose p}}{n!}\left(   \frac{N_im_i}{(1-\rho_i)\overline{\gamma}_i}\right)^n\Gamma(M_{r,i}m_{r,i}+p) \right.\\&~\times\left.x^n\left(\frac{M_{r,i}m_{r,i}}{\overline{\gamma}_{r,i}} + \frac{N_im_ix}{(1-\rho_i)\overline{\gamma}_{i}}\right)^{-(M_{r,i}m_{r,i}+p)}e^{-\frac{N_im_ix}{(1-\rho_i)\overline{\gamma}_i}} \right].
\end{split}
\end{equation}
Differentiating (\ref{eq12}) and after some mathematical manipulations, the PDF of the effective SINR is given by 
\begin{equation}\label{eq13}
\begin{split}
f_{\gamma_{\text{eff},i}}(x) = & \left(\frac{N_im_i}{\overline{\gamma}_i}\right)^{N_im_i}\left(\frac{M_{r,i}m_{r,i}}{\overline{\gamma}_{r,i}}\right)^{M_{r,i}m_{r,i}}\frac{x^{N_im_i-1}}{\Gamma(N_im_i)}\\&\times~
\frac{e^{-\frac{N_im_ix}{(1-\rho_i)\overline{\gamma}_{i}}}}{\Gamma(M_{r,i}m_{r,i})}\sum_{p=0}^{N_im_i}{N_im_i \choose p}\Gamma(M_{r,i}m_{r,i}+p)\\&\times~
\left(\frac{M_{r,i}m_{r,i}}{\overline{\gamma}_{r,i}} + \frac{N_im_ix}{(1-\rho_i)\overline{\gamma}_i} \right)^{-(M_{r,i}m_{r,i}+p)}.
\end{split}
\end{equation}
The $n$-th moment of the SINR of hop $i$ is defined as
\begin{equation}
\mathbb{E}[\gamma^n] = \int\limits_{0}^{\infty}\gamma^n f_{\gamma_{\text{eff},i}}(\gamma)d\gamma.   
\end{equation}
After some mathematical manipulations, the $n$-th moment is given by
\begin{equation}{\label{eq15}}
\begin{split}
 \mathbb{E}[\gamma^n] =& \frac{\left(\frac{N_im_i}{\overline{\gamma}_i}\right)^{N_im_i}}{\Gamma(N_im_i)\Gamma(M_{r,i}m_{r,i})}\sum_{p=0}^{N_im_i}{N_im_i \choose p}\left(\frac{\overline{\gamma}_{r,i}}{M_{r,i}m_{r,i}} \right)^{p}\left(\frac{(1-\rho_i)\overline{\gamma}_i}{N_im_i}\right)^{n+N_im_i}\\&\times~G_{2,1}^{1,2} \Bigg(\frac{\overline{\gamma}_{r,i}}{M_{r,i}m_{r,i}} ~\bigg|~\begin{matrix} 1-n-N_im_i,~1-M_{r,i}m_{r,i}-p \\ 0 \end{matrix} \Bigg). 
\end{split}
\end{equation}
\begin{proof}
The proof of ($\ref{eq15}$) is reported in Appendix A.
\end{proof}
\subsection{High SNR Analysis}
Using the identity $\frac{\gamma(s,x)}{x^s} \backsim \frac{1}{s}$ as $x \rightarrow 0$, at high SNR, (\ref{eq8}) can be expressed as
\begin{equation}\label{eq16}
F_{\hat{\gamma_i}}(x) \backsim \frac{1}{(N_im_i)!}\left(\frac{N_im_ix}{\overline{\gamma}_i}\right)^{N_im_i}.
\end{equation}
After considering (\ref{eq11}) and repeating the same derivation steps, the high SNR asymptote of the CDF of the effective SINR is given by
\begin{equation}\label{eq17}
\begin{split}
F_{\gamma_{\text{eff},i}}(x) \backsim  \frac{1}{\Gamma(M_{r,i}m_{r,i})(N_im_i)!}
\left(\frac{N_im_ix}{\overline{\gamma}_i}\right)^{N_im_i}
\sum_{p=0}^{N_im_i}{N_im_i \choose p}\Gamma(M_{r,i}m_{r,i}+p)\left(\frac{\overline{\gamma}_{r,i}}{M_{r,i}m_{r,i}} \right)^p.
\end{split}
\end{equation}
From (\ref{eq17}), the diversity gain per hop $G_{d,i}$ is
\begin{equation}\label{eq18}
G_{d,i} = N_im_i.
\end{equation}
\subsection{Symbol error probability}
For M-QAM modulations, the average symbol error is expressed as
\begin{equation}\label{eq19}
P_{e,i} = \alpha \mathbb{E}_{\gamma_{\text{eff},i}}\left[\mathcal{Q}(\sqrt{\beta\gamma})\right]
\end{equation}
where $\mathcal{Q}(\cdot)$ is the Gaussian $Q$-function, and $\alpha,~\beta$ stand for the modulation schemes. Using the integration by part, the symbol error can be reformulated as follows
\begin{equation}\label{eq20}
P_{e,i} = \frac{\alpha\sqrt{\beta}}{2\sqrt{2\pi}}\int\limits_0^{\infty}\frac{e^{-\frac{\beta\gamma}{2}}}{\sqrt{\gamma}}F_{\gamma_{\text{eff},i}}(\gamma)d\gamma.
\end{equation}
Inserting (\ref{eq12}) in (\ref{eq20}) and after using \cite[Eq.~(7.813.1)]{15}, and \cite[Eq.~(07.43.03.0271.01)]{18}, the probability of symbol error is given by
\begin{equation}\label{eq21}
\begin{split}
&P_{e,i} = \frac{\alpha\sqrt{\beta}(1-\rho)^{N_im_i}}{2\sqrt{2\pi}}\left(\Gamma(0.5)\sqrt{\frac{2}{\beta}}  - \sum_{n=0}^{N_im_i-1}\sum_{p=0}^{n}\frac{{n \choose p}}{n!} \right.\\&\times~\left.
\frac{\left(\frac{\overline{\gamma}_{r,i}}{m_{r,i}}\right)^p}{\Gamma(M_{r,i}m_{r,i})}\left(\frac{N_im_i}{(1-\rho_i)\overline{\gamma}_i} \right)^n \left(\frac{\beta}{2} + \frac{N_im_i}{(1-\rho_i\overline{\gamma}_i)} \right)^{-(n+\frac{1}{2})}  \right.\\&\times~\left. G_{2,1}^{1,2} \Bigg(\frac{N_im_i\overline{\gamma}_{r,i}}{(1-\rho_i)M_{r,i}m_{r,i}\overline{\gamma}_i}\left(\frac{\beta}{2} + \frac{N_im_i}{(1-\rho_i)\overline{\gamma}_i} \right) ~\bigg|~\begin{matrix} \kappa_1 \\ \kappa_2 \end{matrix} \Bigg)  \right)
\end{split}
\end{equation}
where $G^{m,n}_{p,q}(\cdot|\begin{matrix} - \end{matrix})$ is the Meijer-$G$ function, $\kappa_1 = \left[\frac{1}{2}-n,~1-M_{r,i}m_{r,i}-p\right]$, and $\kappa_2 = 0$.
\begin{proof}
The proof of ($\ref{eq21}$) is reported in Appendix B.
\end{proof}
At high SNR, the probability of error can be expressed as 

\begin{equation}\label{eq22}
P_{e,i} \cong (G_{c,i}\overline{\gamma}_i)^{-G_{d,i}}
\end{equation}
where $G_{c,i}$ is the coding gain. After inserting (\ref{eq17}) in (\ref{eq20}), the high SNR error is derived as follows
\begin{equation}\label{eq23}
\begin{split}
P_{e,i} = \frac{\alpha\Gamma(N_im_i+\frac{1}{2})}{2\sqrt{\pi}(N_im_i)!}\left(\frac{2N_im_i}{\beta\overline{\gamma}_i}\right)^{N_im_i}\sum_{p=0}^{N_im_i}{N_im_i \choose p} 
\frac{\Gamma(M_{r,i}m_{r,i}+p}{\Gamma(M_{r,i}m_{r,i})})\left(\frac{\overline{\gamma}_{r,i}}{m_{r,i}}\right)^p.
\end{split}
\end{equation}
After identifying (\ref{eq23}) with (\ref{eq22}), the coding gain is given by
\begin{equation}\label{eq24}
\begin{split}
G_{c,i} = \frac{\beta}{2N_im_i}\left( \frac{2\Gamma(M_{r,i}m_{r,i})(N_im_i)!\sqrt{\pi}}{\alpha\Gamma(N_im_i+\frac{1}{2})} \right)^{\frac{1}{m}}
\left( \sum_{p=0}^{N_im_i} {N_im_i \choose p}\Gamma(M_{r,i}m_{r,i}+p)\left( \frac{\overline{\gamma}_{r,i}}{M_{r,i}m_{r,i}} \right)^p   \right)^{\frac{1}{m}}.
\end{split}
\end{equation}

\subsection{Capacity}
The average rate $\mathcal{C}_i$, expressed in (bps), is defined as the maximum error-free data rate transmitted by the system. Assuming Gaussian signaling, the mean rate can be written as follows
\begin{equation}\label{eq25}
\mathcal{C}_i = B\mathbb{E}_{\gamma_{\text{eff},i}}\left[ \log_2(1+\gamma)\right].
\end{equation} 
The derivation of the closed-form expression involves three main steps. The first one is to insert (\ref{eq13}) in (\ref{eq25}). Secondly, we convert each function of the integral into Fox-$H$ function form. Finally, after applying the identity \cite[Eq.~(2.3)]{16}, the capacity is given by (\ref{eq26}). Note that the function $H_{p1:q1,p2,q2:p3,q3}^{m1,n1:m2,n2:m3,n3}(-|-|-|\cdot,\cdot)$ is called the bivariate Fox-$H$ function. An efficient MATLAB implementation of this function is given by \cite[Appendix B]{17}.
\begin{equation}\label{eq26}
\begin{split}
\mathcal{C}_i = \frac{B(1-\rho_i)^{N_im_i}}{\log(2)\Gamma(N_im_i)\Gamma(M_{r,i}m_{r,i})}\sum_{p=0}^{N_im_i}{N_im_i \choose p} \left( \frac{\overline{\gamma}_{r,i}}{M_{r,i}m_{r,i}} \right)^p~~~~~~~~~~~~~~~~~~~~~~~~~\\\times~
H_{1,0:1,1:2,2}^{0,1:1,1:1,2} \Bigg( \begin{matrix}  (1-N_im_i;1,1) \\ - \end{matrix} ~\Bigg|~ \begin{matrix}  (1-M_{r,i}m_{r,i}-p,1) \\ (0,1)\end{matrix}~\Bigg|~ \begin{matrix}  (1,1),~(1,1)\\(1,1),~(0,1) \end{matrix} ~\Bigg|~ \frac{\overline{\gamma}_{r,i}}{M_{r,i}m_{r,i}},~\frac{(1-\rho_i)\overline{\gamma}_i}{N_im_i} \Bigg).
\end{split}
\end{equation}
\begin{proof}
The proof of (\ref{eq26}) is reported in Appendix C.
\end{proof}
\subsubsection{Low power regime}
Using the following approximation $\log_2(1+x)\backsim x\log_2(e)$ for $x$ very close to zero, the capacity can be reduced for low power regime as follows
\begin{equation}{\label{eq27}}
\mathcal{C}_i^{\text{low}} \cong B\log_2(e)\mathbb{E}[\gamma^{n=1}].    
\end{equation}
Note that the approximation to the first moment is valid for low power, however, it substantially deviates from the exact expression as the power increases.
\subsubsection{High power regime}
For large transmitted power, the capacity can be approximated as 
\begin{equation}{\label{eq28}}
\mathcal{C}_i^{\text{high}} \cong B\log_2(e)\frac{\partial \mathbb{E}[\gamma^n]}{\partial n}\Bigr\rvert_{n = 0}.
\end{equation}
Note that the deriving the $n$-th with respect to $n$ is not tractable as the parameter $n$ is not the main argument of the Meijer-G function. To solve this problem, we need to interchange the derivative operator (with respect to $n$) with the integral operator (with respect to $\gamma$) before evaluating the $n$-th moment. This approach can be reformulated as follows
\begin{equation}{\label{eq29}}
\frac{\partial \mathbb{E}[\gamma^n]}{\partial n} = \frac{\partial}{\partial n}\Bigg(\int\limits_{0}^{\infty}\gamma^n f_{\gamma_{\text{eff},i}}(\gamma)d\gamma \Bigg) = \int\limits_{0}^{\infty}\frac{\partial \gamma^n}{\partial n} f_{\gamma_{\text{eff},i}}(\gamma)d\gamma = \int\limits_{0}^{\infty}\gamma^n\log(\gamma) f_{\gamma_{\text{eff},i}}(\gamma)d\gamma.  
\end{equation}
Note that a closed-form expression of this integral is not tractable. Hence, a numerical integration is required.
\subsubsection{Jensen's upper bound}
The capacity can be further characterized by a tight upper bound derived from Jensen's inequality as follows
\begin{equation}{\label{eq30}}
\mathcal{C}_i^{\text{ub}} = B \log_2(1+\mathbb{E}[\gamma^{n=1}]).  
\end{equation}
\subsection{Massive MIMO}
For large number of antennas of the relay, the distribution of the SINR of hop $i$ can be approximated to a Gaussian distribution using the central limit theorem. The approximated CDF can be obtained by
\begin{equation}{\label{eq31}}
F_{\gamma_{\text{eff},i}}(x)\cong \frac{1}{2}\left(1 + \erf\left(\frac{x-\mu_i}{\sqrt{2\sigma_i^2}}\right)\right)   
\end{equation}
where $\erf(\cdot)$ is the error function, $\mu_i$ and $\sigma_i^2$ are the mean and the variance of the Gaussian distribution which is given by
\begin{equation}{\label{eq32}}
\mu_i = (1-\rho_i)\overline{\gamma}_i,    
\end{equation}
\begin{equation}{\label{eq33}}
\sigma_i^2 = \frac{(1-\rho_i)^2\overline{\gamma}_i^2}{N_im_i}.     
\end{equation}
\subsection{Blockage model}
mmWaves communications are very sensitive to blockage. Consequently, it is important to consider this factor in the analysis to get consistent results. Blockage models have been extensively studied in the literature \cite{e5,e6,e7}. The proposed models basically depend on the geometry of the objects, and the density $\delta$ in a given area (urban, suburban, and rural). In this work, we will focus on the following blockage models.
\begin{equation}{\label{eq34}}
P_{\text{los}} = e^{-\frac{d}{\delta}}  
\end{equation}
where $P_{\text{los}}$ is the probability of LOS, and $d$ is the distance between the TX and RX. This model is called the exponential blockage model. According to 3GPP, $\delta = 200$ m in suburban area, and $\delta = 63$ m in urban area \cite{10}. The next blockage model is defined as 
\begin{equation}{\label{eq35}}
P_{\text{los}} = \min\left(\frac{18}{d},~1\right)\left(1 - e^{-\frac{d}{63}}\right) +  e^{-\frac{d}{63}}.   
\end{equation}
Note that many external factors related to the environment are subject to the blockage such as the rain attenuation. Generally, these parameters are more severe for NLOS than LOS. Since the blockage causes relative scattering, the Nakagami-$m$ parameter can take 4, and 2 for LOS, and NLOS, respectively. In case of severe scattering, the Nakagami-$m$ parameter can take 1, and 0.5, respectively, for Rayleigh and worst Rayleigh scenarios.
TABLE (\ref{table2}) summarizes the mmWaves propagation parameters for different carrier frequencies.
\begin{table}[H]\label{mmwave}
\renewcommand{\arraystretch}{1}
\caption{mmWaves-Band propagation measurements in different bands. The rain attenuations are measured at 5 mm/h while the oxygen absorption is evaluated at 200 m, \cite[TABLE II]{e5}.}
\label{table2}
\centering
\begin{tabular}{|c|c|c|c|}
\hline
\bfseries Frequency [GHz] & \bfseries $\alpha_{\text{rain,~los}}$ [dB]  &  \bfseries \bfseries $\alpha_{\text{rain,~nlos}}$ [dB]&\bfseries $\alpha_{\text{ox}}$ [dB] \\
\hline
28 & 0.18   & 0.9 & 0.04\\
\hline
38 & 0.26   & 1.4 & 0.03\\
\hline
60 & 0.44& 2.0 &3.2\\
\hline
73& 0.6 & 2.4&0.09\\
\hline
\end{tabular}
\end{table}
\section{Relay Channel Reliability Metrics Analysis}
In this section, we provide the performance analysis of the proposed system, in particular, we consider the end-to-end outage probability, the probability of symbol error, and the ergodic capacity for the overall relay system. Basically, the statistical results derived in the previous section will be considered to get the reliability metrics and to quantify the performance losses and the system gains.
\subsection{Outage probability}
The outage probability is defined as the probability that the overall SINR falls below a given threshold $x$. This threshold is a standard value designed to target a desirable quality of service.

Since LOS and NLOS propagations are assumed, the average CDF per hop is given by
\begin{equation}\label{eq36}
F_{\gamma_{\text{eff},i}}(x) = P_{\text{los}}F^{\text{los}}_{\gamma_{\text{eff},i}}(x) + (1 - P_{\text{los}})F^{\text{nlos}}_{\gamma_{\text{eff},i}}(x)
\end{equation}
where $F^{\text{los}}_{\gamma_{\text{eff},i}}(\cdot)$ and $F^{\text{nlos}}_{\gamma_{\text{eff},i}}(\cdot)$ are the CDF of hop $i$ evaluated respectively, for LOS and NLOS.

The overall SINR $\gamma_e$ for the relay system employing DF protocol is given by
\begin{equation}\label{eq37}
\gamma_e = \min\left(\gamma_{\text{eff},1},~ \gamma_{\text{eff},2}\right).
\end{equation}
Consequently, the CDF of $\gamma_e$ is given by
\begin{equation}\label{eq38}
F_{\gamma_e}(x) = F_{\gamma_{\text{eff},1}}(x) + F_{\gamma_{\text{eff},2}}(x) - F_{\gamma_{\text{eff},1}}(x)F_{\gamma_{\text{eff},2}}(x).
\end{equation}
Finally, the outage probability is obtained by
\begin{equation}\label{eq39}
P_{\text{out}}(x) = \mathbb{P}[\gamma_e < x] = F_{\gamma_e}(x).
\end{equation}
\subsection{Symbol error probability}
After plugging (\ref{eq38}) in (\ref{eq20}), the symbol error probability of the overall relay channel can be obtained by
%\newpage
%\begin{strip}
%\noindent\rule{18cm}{1pt}
\begin{equation}\label{eq40}
\begin{split}
P_e =& (1-(1-\rho_2)^{N_2m_2})P_{e,1} + (1-(1-\rho_1)^{N_1m_1})P_{e,2}+\frac{(1-\rho_1)^{N_1m_1}(1-\rho_2)^{N_2m_2}\alpha\Gamma(0.5)}{2\sqrt{\pi}}\\&+\frac{\alpha\sqrt{\beta}(1-\rho_1)^{N_1m_1}(1-\rho_2)^{N_2m_2}}{2\sqrt{2\pi}\Gamma(M_{r,1}m_{r,1})\Gamma(M_{r,2}m_{r,2})}\sum_{n_1=0}^{N_1m_1-1}\sum_{n_2=0}^{N_2m_2-1}\sum_{p_1=0}^{n_1}\sum_{p_2=0}^{n_2}\frac{{n_1\choose p_1}{n_2\choose p_2}}{n_1!n_2!\lambda^{n_1+n_2+\frac{1}{2}}}
\left(\frac{N_1m_1}{(1-\rho_1)\overline{\gamma}_1}\right)^{n_1}\\~&\times\left(\frac{N_2m_2}{(1-\rho_2)\overline{\gamma}_2}\right)^{n_2}\left(\frac{\overline{\gamma}_{r,1}}{M_{r,1}m_{r,1}}\right)^{p_1}\left(\frac{\overline{\gamma}_{r,2}}{M_{r,2}m_{r,2}}\right)^{p_2}
\\& H_{1,0:1,1:1,1}^{0,1:1,1:1,1}\Bigg( \begin{matrix}  (\frac{1}{2}-n_1-n_2;1,1) \\ - \end{matrix} ~\Bigg|~ \begin{matrix}  (1-M_{r,1}m_{r,1}-p_1,1) \\ (0,1)\end{matrix}~\Bigg|~ \begin{matrix}  (1-M_{r,2}m_{r,2}-p_2,1) \\ (0,1) \end{matrix} ~\Bigg|~ \nu_1,~\nu_2 \Bigg)  
\end{split}
\end{equation}
%\noindent\rule{18cm}{1pt}
%\end{strip}
where $\lambda$, and $\nu_i$ are given by
\begin{equation}\label{eq41}
\lambda = \frac{\beta}{2} + \frac{N_1m_1\overline{\gamma}_{r,1}}{M_{r,1}m_{r,1}(1-\rho_1)\overline{\gamma}_1} + \frac{N_2m_2\overline{\gamma}_{r,2}}{M_{r,2}m_{r,2}(1-\rho_2)\overline{\gamma}_2}.
\end{equation}
\begin{equation}{\label{eq42}}
\nu_i = \frac{N_im_i\overline{\gamma}_{r,i}}{M_{r,i}m_{r,i}(1-\rho_i)\overline{\gamma}_i\lambda},~~i = 1,~2.
\end{equation}
\begin{proof}
The proof of (\ref{eq40}) is reported in Appendix D.
\end{proof}
\subsection{Ergodic capacity}
Since DF is assumed, the overall capacity of the relay channel can be expressed as follows
\begin{equation}\label{eq43}
\mathcal{C} = \min\left(\mathcal{C}_1,~\mathcal{C}_2\right).
\end{equation}
\subsection{Capacity with outage}
For slower vehicle, the channel fading can be viewed as slow fading where the receive SINR of the channel is randomly distributed taking realizations closer to zero. In this case, the capacity is zero most of the time and an alternative reliable metric is the outage capacity.

For a given outage $\epsilon$, the $\epsilon$-capacity\footnote[1]{In this work, we define the notion of capacity with outage as the average data rate that is correctly received/decoded at the receiver which is equivalent to the throughput. In other standards in the literature, the capacity with outage is assimilated with the transmit data rate. The only difference is if we consider capacity with outage as the throughput, we account for the probability of bursts (outage) and we multiply by the term ($1-\epsilon$), while for the transmit data rate, the term (1-$\epsilon$) is not accounted anymore.} is defined as the maximum data rate maintained with a probability of (1-$\epsilon$). It can be expressed as follows
\begin{equation}\label{eq44}
\mathcal{C}_{\epsilon} = (1-\epsilon)B\mathbb{E}_{\gamma_e}[\log_2(1+\gamma_{\epsilon})]
\end{equation}
where $\gamma_{\epsilon}$ is the minimum receive SINR for which the channel can support the rate $\mathcal{C}_{\epsilon}$. In other terms, if the receive SINR is less than $\gamma_{\epsilon}$, the system is in outage, otherwise, the received data can be successfully decoded.
\subsection{System gains}
The overall diversity gain $G_d$ is given by
\begin{equation}\label{eq45}
G_d = \min\left(G_{d,1},~G_{d,2}\right).
\end{equation}
while the overall coding gain $G_c$ is obtained by
\begin{equation}\label{eq46}
G_c = \left\{
        \begin{array}{ll}
            G_{c,1} & \quad G_{d,1} < G_{d,2} \\
            \left(G_{c,1}^{-G_d} + G_{c,2}^{-G_d}\right)^{-\frac{1}{G_d}} & \quad G_{d,1} = G_{d,2}\\
            G_{c,2}& \quad G_{d,1} > G_{d,2}
        \end{array}
    \right.
\end{equation}
\subsection{Block fading channel model}
Since the channel is time-varying, it can be subdivided into a set of time block-fading narrowband sub-channels where the duration of each fading block is the coherence time. This means that for each block, a new channel $\boldsymbol{H}[k]$ is drawn from a Nakagami-m distribution, and it is completely uncorrelated with $\boldsymbol{H}[k-1]$. Thereby, all the reliability metrics can be obtained by averaging over the total number of blocks $K$. 
\section{Numerical Results}
\begin{figure}[H]
\centering
\includegraphics[scale=1]{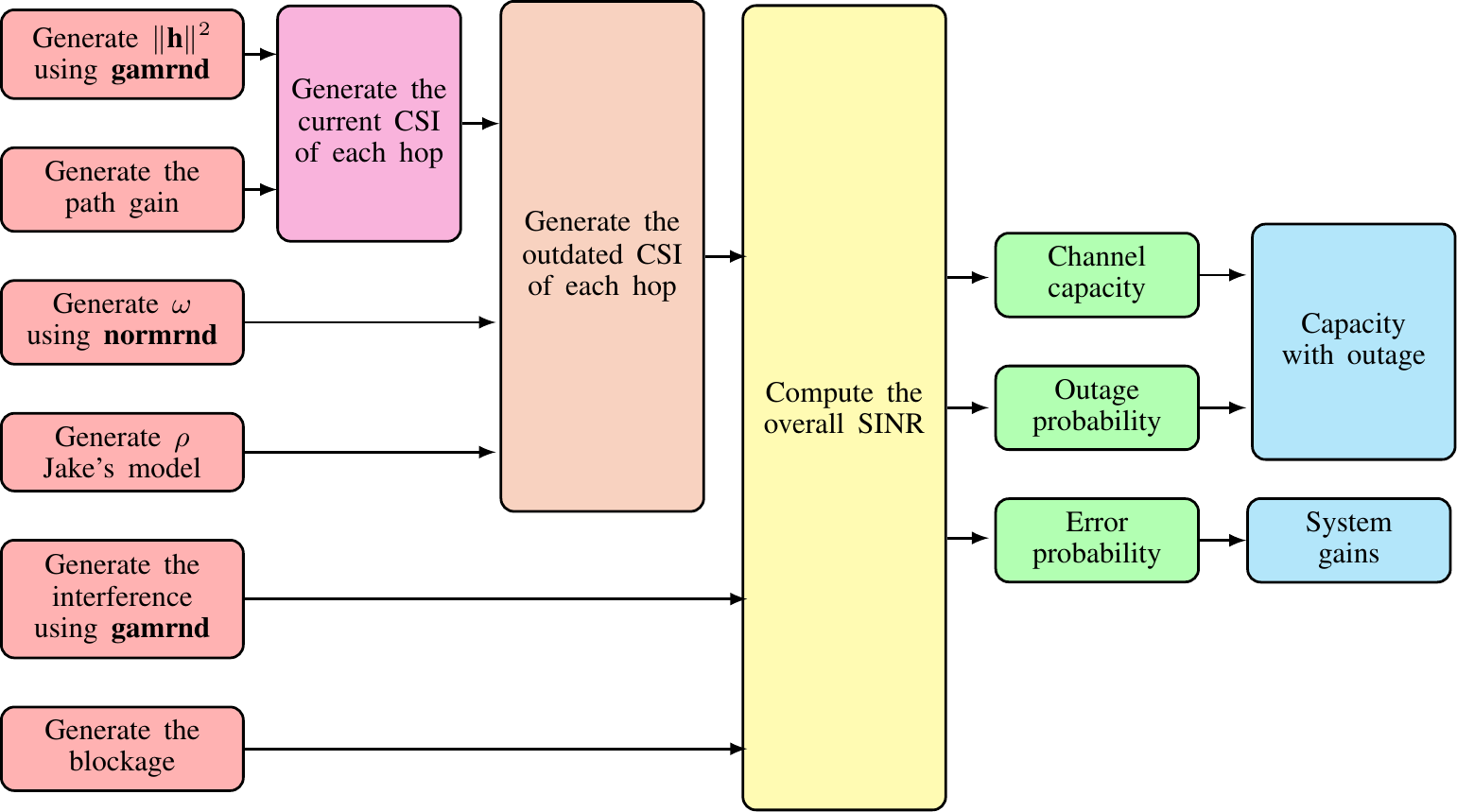}
    \caption{mBlock diagram of the simulation setup.}
    \label{sim}
\end{figure}

In this section, we present the analytical results of the reliability metrics following their discussion. We also check the accuracy of these derived expressions with Monte Carlo simulations \footnote[2]{For all cases, $10^6$ realizations of the random variables were generated to perform the Monte Carlo simulation in MATLAB.}. Unless otherwise stated, TABLE (\ref{table3}) provides the signal and the interference parameters values. Fig.~\ref{sim} presents the block diagram explaining the detailed steps of the Monte Carlo simulation to get the performance metrics. 

\begin{figure}[H]
\centering
\setlength\fheight{5cm}
\setlength\fwidth{7cm}
\definecolor{mycolor1}{rgb}{0.00000,0.44700,0.74100}%
\definecolor{mycolor2}{rgb}{0.85000,0.33000,0.10000}%
\begin{tikzpicture}

\begin{axis}[%
width=0.951\fwidth,
height=\fheight,
at={(0\fwidth,0\fheight)},
scale only axis,
xmin=-12,
xmax=10,
axis x line*=bottom,
axis y line*=left,
xlabel style={font=\color{white!15!black}},
xlabel={\sffamily{Average SNR (dB)}},
ymode=log,
ymin=1e-09,
ymax=1,
yminorticks=true,
ylabel style={font=\color{white!15!black}},
ylabel={\sffamily{Outage Probability}},
axis background/.style={fill=white},
legend style={legend cell align=left, align=left, draw=white!15!black,draw=none,fill=none}
]
\addplot [color=black, line width=1pt]
  table[row sep=crcr]{%
-11.5204205106835	0.000473999499562719\\
-10.5204205106835	0.000412307986314616\\
-9.52042051068354	0.000331912559228692\\
-8.52042051068354	0.00024507301593261\\
-7.52042051068354	0.000165365204361654\\
-6.52042051068354	0.000102000099161398\\
-5.52042051068354	5.76956649722652e-05\\
-4.52042051068354	3.0084625750833e-05\\
-3.52042051068354	1.45576996438084e-05\\
-2.52042051068354	6.5864041249231e-06\\
-1.52042051068354	2.80821656998091e-06\\
-0.520420510683537	1.13719727340594e-06\\
0.479579489316461	4.40653973654717e-07\\
1.47957948931646	1.64509629755834e-07\\
2.47957948931646	5.9535830387429e-08\\
3.47957948931646	2.09984130843791e-08\\
4.47957948931646	7.25125473072854e-09\\
5.47957948931646	2.46120162434765e-09\\
6.47957948931646	8.2375612756179e-10\\
7.47957948931646	2.72604697668645e-10\\
8.47957948931646	8.93937855698344e-11\\
9.47957948931646	2.91004525832411e-11\\
10.4795794893165	9.41764204069774e-12\\
11.4795794893165	3.03350377429794e-12\\
12.4795794893165	9.73458030323176e-13\\
13.4795794893165	3.11450572082635e-13\\
14.4795794893165	9.94089056224149e-14\\
15.4795794893165	3.16688279549381e-14\\
16.4795794893165	1.00737354231026e-14\\
17.4795794893165	3.20030735259497e-15\\
18.4795794893165	1.01574609249867e-15\\
};
\addlegendentry{\sffamily{$\rho{}_\text{1}\text{ = 0.01}$}}

\addplot [color=mycolor1, dash pattern={on 7pt off 2pt on 0pt off 0pt}, line width=1pt]
  table[row sep=crcr]{%
-11.5204205106835	0.00020332198684505\\
-10.5204205106835	0.000186103535117844\\
-9.52042051068354	0.000159457610112939\\
-8.52042051068354	0.000126144939998868\\
-7.52042051068354	9.14244792666468e-05\\
-6.52042051068354	6.05389796332541e-05\\
-5.52042051068354	3.66617274536756e-05\\
-4.52042051068354	2.0379162921265e-05\\
-3.52042051068354	1.04562888913931e-05\\
-2.52042051068354	4.98621706001917e-06\\
-1.52042051068354	2.22678147242511e-06\\
-0.520420510683537	9.38695804964861e-07\\
0.479579489316461	3.7643119106684e-07\\
1.47957948931646	1.446592061217e-07\\
2.47957948931646	5.36314339712132e-08\\
3.47957948931646	1.92973854611215e-08\\
4.47957948931646	6.77393583889132e-09\\
5.47957948931646	2.33012081336807e-09\\
6.47957948931646	7.88385382676869e-10\\
7.47957948931646	2.63195097457244e-10\\
8.47957948931646	8.69191450497709e-11\\
9.47957948931646	2.84556276682974e-11\\
10.4795794893165	9.25085876518996e-12\\
11.4795794893165	2.99062080095477e-12\\
12.4795794893165	9.6248408543671e-13\\
13.4795794893165	3.08653129173974e-13\\
14.4795794893165	9.86976851527434e-14\\
15.4795794893165	3.1488739549328e-14\\
16.4795794893165	1.00280527984947e-14\\
17.4795794893165	3.18902060494978e-15\\
18.4795794893165	1.01287551780769e-15\\
};
\addlegendentry{\sffamily{$\rho{}_\text{1}\text{ = 0.75}$}}

\addplot [color=mycolor2, dash pattern={on 7pt off 2pt on 1pt off 3pt}, line width=1pt]
  table[row sep=crcr]{%
-11.5204205106835	1.67484689293924e-05\\
-10.5204205106835	1.656065183089e-05\\
-9.52042051068354	1.6006866367637e-05\\
-8.52042051068354	1.48284973811459e-05\\
-7.52042051068354	1.29079725274093e-05\\
-6.52042051068354	1.03999371508736e-05\\
-5.52042051068354	7.68600470445655e-06\\
-4.52042051068354	5.19103652703117e-06\\
-3.52042051068354	3.20484436979406e-06\\
-2.52042051068354	1.81437332099465e-06\\
-1.52042051068354	9.46843526186695e-07\\
-0.520420510683537	4.58505394889351e-07\\
0.479579489316461	2.07579006277236e-07\\
1.47957948931646	8.85551110187194e-08\\
2.47957948931646	3.58783504991067e-08\\
3.47957948931646	1.39083578822474e-08\\
4.47957948931646	5.19423855097244e-09\\
5.47957948931646	1.88033685447317e-09\\
6.47957948931646	6.63358237887165e-10\\
7.47957948931646	2.29118558078019e-10\\
8.47957948931646	7.77792652175536e-11\\
9.47957948931646	2.60358415568872e-11\\
10.4795794893165	8.61691354564847e-12\\
11.4795794893165	2.82593569515298e-12\\
12.4795794893165	9.19991914315801e-13\\
13.4795794893165	2.97748739658507e-13\\
14.4795794893165	9.59115812312974e-14\\
15.4795794893165	3.07792967138428e-14\\
16.4795794893165	9.84788021250045e-15\\
17.4795794893165	3.14331299254168e-15\\
18.4795794893165	1.0013940018315e-15\\
};
\addlegendentry{\sffamily{$\rho{}_\text{1}\text{ = 0.85}$}}

\addplot [color=black, dotted, line width=1pt]
  table[row sep=crcr]{%
-11.5204205106835	2.2118748276523e-06\\
-10.5204205106835	2.21057229202778e-06\\
-9.52042051068354	2.20224072193952e-06\\
-8.52042051068354	2.16912069281537e-06\\
-7.52042051068354	2.07935914816515e-06\\
-6.52042051068354	1.90101936242357e-06\\
-5.52042051068354	1.62635062389331e-06\\
-4.52042051068354	1.28434233125167e-06\\
-3.52042051068354	9.29118420823295e-07\\
-2.52042051068354	6.14094982939489e-07\\
-1.52042051068354	3.71216751208228e-07\\
-0.520420510683537	2.05995474019656e-07\\
0.479579489316461	1.05526679567786e-07\\
1.47957948931646	5.02498668147253e-08\\
2.47957948931646	2.24125081244262e-08\\
3.47957948931646	9.43750994813064e-09\\
4.47957948931646	3.78099166925447e-09\\
5.47957948931646	1.45183332080366e-09\\
6.47957948931646	5.37894864437601e-10\\
7.47957948931646	1.93434378498624e-10\\
8.47957948931646	6.78697873317154e-11\\
9.47957948931646	2.33372976148648e-11\\
10.4795794893165	7.89365188340566e-12\\
11.4795794893165	2.63457041696605e-12\\
12.4795794893165	8.69883078805488e-13\\
13.4795794893165	2.84737070407193e-13\\
14.4795794893165	9.25554687839833e-14\\
15.4795794893165	2.99182864510499e-14\\
16.4795794893165	9.62793738227061e-15\\
17.4795794893165	3.08732167519466e-15\\
18.4795794893165	9.87177717160735e-16\\
};
\addlegendentry{\sffamily{$\rho{}_\text{1}\text{ = 0.90}$}}

\addplot [color=black, dash pattern={on 7pt off 2pt on 1pt off 3pt} , line width=1pt]
  table[row sep=crcr]{%
-11.5204205106835	1.03053850469823\\
-10.5204205106835	0.325884889135053\\
-9.52042051068354	0.103053850469823\\
-8.52042051068354	0.0325884889135053\\
-7.52042051068354	0.0103053850469823\\
-6.52042051068354	0.00325884889135053\\
-5.52042051068354	0.00103053850469823\\
-4.52042051068354	0.000325884889135053\\
-3.52042051068354	0.000103053850469823\\
-2.52042051068354	3.25884889135053e-05\\
-1.52042051068354	1.03053850469823e-05\\
-0.520420510683537	3.25884889135052e-06\\
0.479579489316461	1.03053850469823e-06\\
1.47957948931646	3.25884889135053e-07\\
2.47957948931646	1.03053850469823e-07\\
3.47957948931646	3.25884889135053e-08\\
4.47957948931646	1.03053850469823e-08\\
5.47957948931646	3.25884889135053e-09\\
6.47957948931646	1.03053850469823e-09\\
7.47957948931646	3.25884889135053e-10\\
8.47957948931646	1.03053850469823e-10\\
9.47957948931646	3.25884889135052e-11\\
10.4795794893165	1.03053850469822e-11\\
11.4795794893165	3.25884889135054e-12\\
12.4795794893165	1.03053850469823e-12\\
13.4795794893165	3.25884889135053e-13\\
14.4795794893165	1.03053850469822e-13\\
15.4795794893165	3.25884889135052e-14\\
16.4795794893165	1.03053850469823e-14\\
17.4795794893165	3.25884889135053e-15\\
18.4795794893165	1.03053850469823e-15\\
};
\addlegendentry{\sffamily{High SNR}}

\end{axis}
\end{tikzpicture}%
    \caption{Probability of outage versus the average SNR for various time correlation values. The carrier frequency is 60 GHz and the link range is 50 m. The number of transmit and receive antennas are 5 and the time correlation for the second hop $\rho_2$ = 0.7. Two interferers are assumed with an average receive SNR of 0 dB.}
     \label{fig3}
\end{figure}
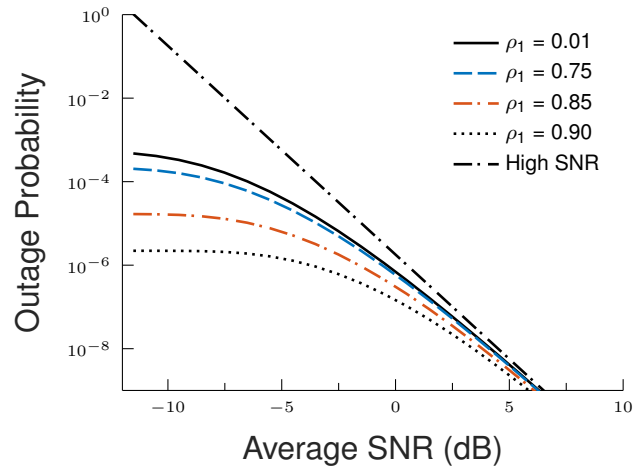

\begin{table}[H]
\renewcommand{\arraystretch}{1}
\caption{Nakagami-$m$ parameters values of the signal and the interference for LOS and NLOS propagation.}
\label{table3}
\centering
\begin{tabular}{|c|c|c|}
\hline
\bfseries Parameter & \bfseries Symbol & \bfseries Value \\
\hline
 Signal (LOS) & $m_{\text{los}}$ & 4 \\
\hline
 Signal (NLOS) & $m_{\text{nlos}}$ & 2\\
\hline
 Interference (LOS) & $m_{\text{r,los}}$ & 3\\
\hline
Interference (NLOS) & $m_{\text{r,nlos}}$ & 1\\
\hline
Number of interferers& $M_r$& 7\\
\hline
\end{tabular}
\end{table}
%\begin{figure}[H]
%\centering
%\setlength\fheight{6cm}
%\setlength\fwidth{12cm}
%\input{block_simulation.tikz}
%    \caption{Block diagram of the simulation setup.}
%    \label{sim}
%\end{figure}
Fig.~\ref{fig3} shows the dependence of the outage probability with respect to the time correlation of the first hop. The results show that the outage performance gets much worse when the channel estimation is bad, while the performance improves as the system becomes more or less able to track the channel variations and gets a better version of the estimated channel. We also observe that the impacts of the time correlation disappear at high SINR as all the characteristics for the different correlation values converge to the same asymptote. This result can be obtained from the high SINR asymptote (\ref{eq17}) where the expression is independent of the time correlation.

\begin{figure}[H]
\centering
\setlength\fheight{5cm}
\setlength\fwidth{7cm}
\input{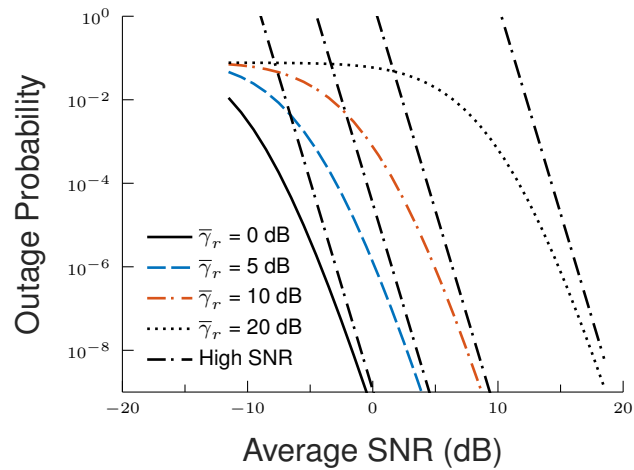}
    \caption{Probability of outage versus the average SNR for different receive SNR values of the interferers. The carrier frequency is 60 GHz and the link range is 50 m. The number of transmit and receive antennas are 5 and the time correlation is $\rho_1 = \rho_2$ = 0.1.}
     \label{fig4}
\end{figure}

Fig.~\ref{fig4} illustrates the impacts of the power interferers on the outage performance. We observe that when the system is saturated by a large received power of the interferers, the outage deteriorates. Inversely, when the receiver becomes less susceptible to interferers (less power), the system performance gets much better. 

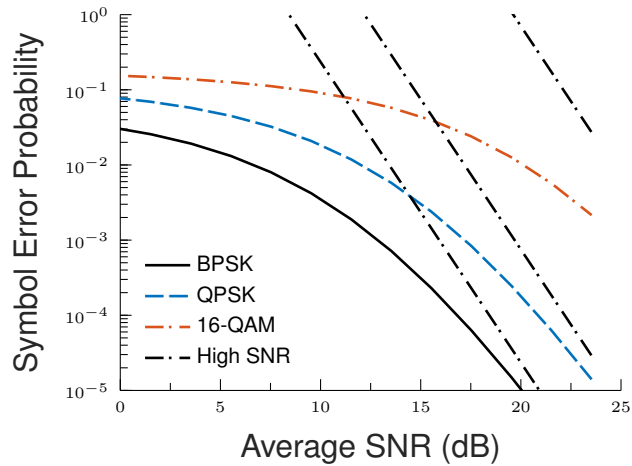
\begin{figure}[H]
\centering
\setlength\fheight{5cm}
\setlength\fwidth{7cm}
\definecolor{mycolor1}{rgb}{0.00000,0.44700,0.74100}%
\definecolor{mycolor2}{rgb}{0.85000,0.33000,0.10000}%
\begin{tikzpicture}

\begin{axis}[%
width=0.951\fwidth,
height=\fheight,
at={(0\fwidth,0\fheight)},
scale only axis,
xmin=0,
xmax=25,
axis x line*=bottom,
axis y line*=left,
xlabel style={font=\color{white!15!black}},
xlabel={\sffamily{Average SNR (dB)}},
ymode=log,
ymin=1e-05,
ymax=1,
yminorticks=true,
ylabel style={font=\color{white!15!black}},
ylabel={\sffamily{Symbol Error Probability}},
axis background/.style={fill=white},
legend style={at={(0.03,0.03)}, anchor=south west, legend cell align=left, align=left, draw=white!15!black,draw=none,fill=none}
]
\addplot [color=black, line width=1pt]
  table[row sep=crcr]{%
-6.46782460524783	0.0460101163018939\\
-4.46782460524783	0.0420241218317782\\
-2.46782460524783	0.0372650431573906\\
-0.467824605247827	0.0317458253420973\\
1.53217539475217	0.025617722349451\\
3.53217539475217	0.0192311986631951\\
5.53217539475217	0.0131401487267303\\
7.53217539475217	0.00797802159906779\\
9.53217539475217	0.00420583591464597\\
11.5321753947522	0.00189026893725225\\
13.5321753947522	0.000717088614008244\\
15.5321753947522	0.000229690653464037\\
17.5321753947522	6.27893411367139e-05\\
19.5321753947522	1.49334654049253e-05\\
21.5321753947522	3.16650890792305e-06\\
23.5321753947522	6.11293250301262e-07\\
};
\addlegendentry{\sffamily{BPSK}}

\addplot [color=mycolor1, dash pattern={on 7pt off 2pt on 0pt off 0pt} , line width=1pt]
  table[row sep=crcr]{%
-6.46782460524783	0.101398865058242\\
-4.46782460524783	0.0954960094290944\\
-2.46782460524783	0.0882629884748941\\
-0.467824605247827	0.0795346866003642\\
1.53217539475217	0.0692470007740129\\
3.53217539475217	0.0575455771917834\\
5.53217539475217	0.044918485637501\\
7.53217539475217	0.0322799999285645\\
9.53217539475217	0.0208669549383631\\
11.5321753947522	0.0118458489915314\\
13.5321753947522	0.00578143020809031\\
15.5321753947522	0.00239082867783199\\
17.5321753947522	0.00083354436080478\\
19.5321753947522	0.00024637674714046\\
21.5321753947522	6.26895066184525e-05\\
23.5321753947522	1.40410451296503e-05\\
};
\addlegendentry{\sffamily{QPSK}}

\addplot [color=mycolor2, dash pattern={on 7pt off 2pt on 1pt off 3pt} , line width=1pt]
  table[row sep=crcr]{%
-6.46782460524783	0.171513162192448\\
-4.46782460524783	0.167402527580283\\
-2.46782460524783	0.162255920807533\\
-0.467824605247827	0.155832805344309\\
1.53217539475217	0.14785679878886\\
3.53217539475217	0.138030348905682\\
5.53217539475217	0.126072365495334\\
7.53217539475217	0.111795129472172\\
9.53217539475217	0.095237476026292\\
11.5321753947522	0.0768531670483528\\
13.5321753947522	0.0576935959895855\\
15.5321753947522	0.039420446180191\\
17.5321753947522	0.0239340647972034\\
19.5321753947522	0.0126175077439379\\
21.5321753947522	0.00567080681175676\\
23.5321753947522	0.00215126584202472\\
};
\addlegendentry{\sffamily{16-QAM}}

\addplot [color=black, dash pattern={on 7pt off 2pt on 1pt off 3pt}  , line width=1pt]
  table[row sep=crcr]{%
-6.46782460524783	906301.479893951\\
-4.46782460524783	143639.104580136\\
-2.46782460524783	22765.2639020267\\
-0.467824605247827	3608.05117829028\\
1.53217539475217	571.837575052356\\
3.53217539475217	90.6301479893951\\
5.53217539475217	14.3639104580136\\
7.53217539475217	2.27652639020267\\
9.53217539475217	0.360805117829028\\
11.5321753947522	0.0571837575052356\\
13.5321753947522	0.00906301479893951\\
15.5321753947522	0.00143639104580135\\
17.5321753947522	0.000227652639020267\\
19.5321753947522	3.60805117829028e-05\\
21.5321753947522	5.71837575052357e-06\\
23.5321753947522	9.06301479893951e-07\\
};
\addlegendentry{\sffamily{High SNR}}

\addplot [color=black, dash pattern={on 7pt off 2pt on 1pt off 3pt}  , line width=1pt, forget plot]
  table[row sep=crcr]{%
-6.46782460524783	29001647.3566064\\
-4.46782460524783	4596451.34656434\\
-2.46782460524783	728488.444864854\\
-0.467824605247827	115457.637705289\\
1.53217539475217	18298.8024016754\\
3.53217539475217	2900.16473566064\\
5.53217539475217	459.645134656434\\
7.53217539475217	72.8488444864854\\
9.53217539475217	11.5457637705289\\
11.5321753947522	1.82988024016754\\
13.5321753947522	0.290016473566064\\
15.5321753947522	0.0459645134656433\\
17.5321753947522	0.00728488444864854\\
19.5321753947522	0.00115457637705289\\
21.5321753947522	0.000182988024016754\\
23.5321753947522	2.90016473566064e-05\\
};
\addplot [color=black, dash pattern={on 7pt off 2pt on 1pt off 3pt}  , line width=1pt, forget plot]
  table[row sep=crcr]{%
-6.46782460524783	27189044396.8185\\
-4.46782460524783	4309173137.40406\\
-2.46782460524783	682957917.0608\\
-0.467824605247827	108241535.348708\\
1.53217539475217	17155127.2515707\\
3.53217539475217	2718904.43968185\\
5.53217539475217	430917.313740407\\
7.53217539475217	68295.7917060801\\
9.53217539475217	10824.1535348708\\
11.5321753947522	1715.51272515707\\
13.5321753947522	271.890443968185\\
15.5321753947522	43.0917313740405\\
17.5321753947522	6.82957917060801\\
19.5321753947522	1.08241535348708\\
21.5321753947522	0.171551272515707\\
23.5321753947522	0.0271890443968185\\
};
\end{axis}
\end{tikzpicture}%
    \caption{Probability of symbol error versus the average SNR for different modulation schemes. The carrier frequency is 38 GHz and the link range is 50 m. The time correlation is $\rho_1 = \rho_2$ = 0.5. We assume only LOS propagation in this scenario.}
     \label{fig5}
\end{figure}

Fig.~\ref{fig5} shows the variations of the symbol error for various modulation schemes. This result is expected as BPSK modulation outperforms QPSK and 16-QAM. Also, this result is considered as a benchmarking tool to verify the accuracy of the symbol error derivation. 
\begin{figure}[H]
\centering
\setlength\fheight{5cm}
\setlength\fwidth{7cm}
\input{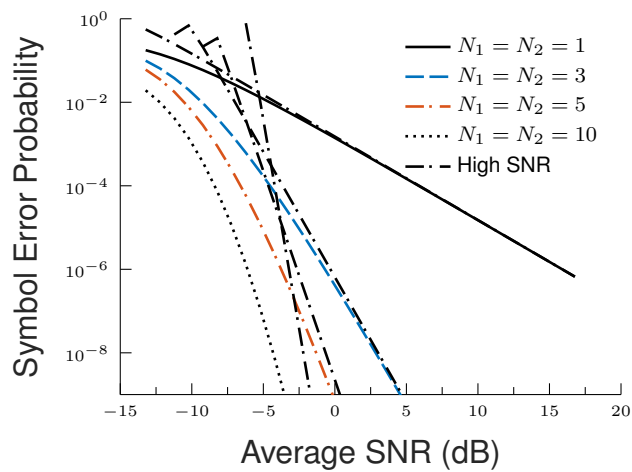}
    \caption{Probability of symbol error versus the average SNR for different number of transmit and receive antennas. The carrier frequency is 73 GHz and the link range is 50 m. The time correlation is $\rho_1 = \rho_2$ = 0.1.}
     \label{fig6}
\end{figure}

Fig.~\ref{fig6} shows the outage performance for various number of transmit and receive antennas. These results show how much diversity gain achieved by the system scaled by the number of antennas. As the number of antennas increase, the system becomes robust enough to combat the fading and improve the error performance. To achieve a given outage of $10^{-3}$, the system setting for 10 antennas achieves an array gain around 15 dB compared to the single antenna configuration.

\begin{figure}[H]
\centering
\setlength\fheight{5cm}
\setlength\fwidth{7cm}
\definecolor{mycolor1}{rgb}{0.00000,0.44700,0.74100}%
\definecolor{mycolor2}{rgb}{0.85000,0.33000,0.10000}%
\begin{tikzpicture}

\begin{axis}[%
width=0.951\fwidth,
height=\fheight,
at={(0\fwidth,0\fheight)},
scale only axis,
xmin=0,
xmax=30,
axis x line*=bottom,
axis y line*=left,
xlabel style={font=\color{white!15!black}},
xlabel={\sffamily{Average Transmit Power (dB)}},
ymin=0,
ymax=4,
ylabel style={font=\color{white!15!black}},
ylabel={\sffamily{Capacity (Gbps)}},
axis background/.style={fill=white},
legend style={at={(0.03,0.97)}, anchor=north west, legend cell align=left, align=left, draw=white!15!black,draw=none,fill=none}
]
\addplot [color=black, line width=1pt]
  table[row sep=crcr]{%
0	0.079091175679441\\
2	0.117030868568131\\
4	0.169012546951394\\
6	0.237320770590483\\
8	0.323212980560514\\
10	0.426615249456362\\
12	0.546178222069809\\
14	0.679658922815107\\
16	0.824453847012488\\
18	0.97809062846589\\
20	1.13856813493346\\
22	1.30453799998095\\
24	1.475384444809\\
26	1.65127250454124\\
28	1.83321382589132\\
30	2.02315980600288\\
};
\addlegendentry{\sffamily{$\delta =$ 50 m}}

\addplot [color=mycolor1, dash pattern={on 7pt off 2pt on 0pt off 0pt},line width=1pt]
  table[row sep=crcr]{%
0	0.130334715498739\\
2	0.192848921526653\\
4	0.278492360412073\\
6	0.391018725882331\\
8	0.532480743239395\\
10	0.702724160276156\\
12	0.899473061042868\\
14	1.11894847503279\\
16	1.35673025501912\\
18	1.6085411956723\\
20	1.87076700329288\\
22	2.14069401384821\\
24	2.41655040745178\\
26	2.6974591348917\\
28	2.98338251645905\\
30	3.27509185941724\\
};
\addlegendentry{\sffamily{$\delta =$ 100 m}}

\addplot [color=mycolor2, dash pattern={on 7pt off 2pt on 1pt off 3pt}, line width=1pt]
  table[row sep=crcr]{%
0	0.153954216656154\\
2	0.227795464952844\\
4	0.328954498466289\\
6	0.461862173577588\\
8	0.628937783097173\\
10	0.829990050994833\\
12	1.06231598499893\\
14	1.32142863396862\\
16	1.60207051213858\\
18	1.89913251614579\\
20	2.20825680356153\\
22	2.52610040862425\\
24	2.85035866157223\\
26	3.17967418899556\\
28	3.51352564891801\\
30	3.85214041400383\\
};
\addlegendentry{\sffamily{$\delta =$ 150 m}}

\addplot [color=black, dotted, line width=1pt]
  table[row sep=crcr]{%
0	0.167324809159153\\
2	0.247578100852977\\
4	0.357520245060863\\
6	0.501965428270727\\
8	0.683540450841444\\
10	0.90203307819468\\
12	1.15449855751399\\
14	1.43604916202424\\
16	1.74095340304566\\
18	2.06363125407698\\
20	2.39930396256953\\
22	2.74427232589957\\
24	3.09593003717517\\
26	3.45264781973667\\
28	3.81363053873471\\
30	4.17879765483209\\
};
\addlegendentry{\sffamily{$\delta =$ 200 m}}

\end{axis}
\end{tikzpicture}%
    \caption{Capacity versus the average transmit power for different blockage densities. The carrier frequency is 28 GHz, the link range is 50 m, and the time correlation is $\rho_1 = \rho_2$ = $10^{-3}$.}
     \label{fig7}
\end{figure}
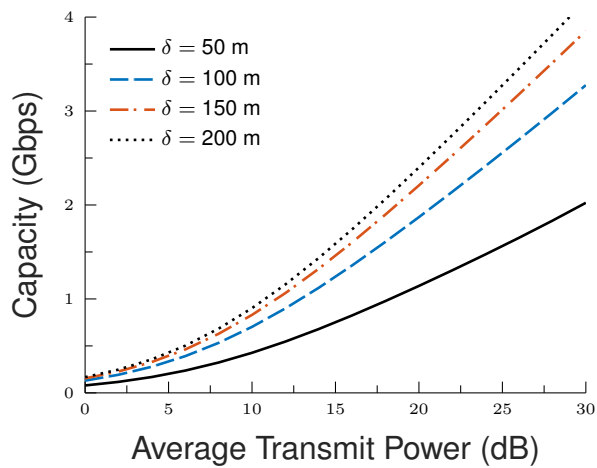

Fig.~\ref{fig7} illustrates the effects of the blockage density on the capacity. As the environment becomes more dense (lower $\delta$) with physical objects, the capacity substantially decreases. This is explained by the fact that mmWaves beams have poor scattering and penetration through the objects. As the blockage becomes severe, the mmWaves beams become more sensitive to the blocking and the signal is completely attenuated. In addition, the system achieves high data rate of Gbps even though for severe blockage constraints. For higher density ($\delta$ = 50 m), the maximum achievable rate is around 2 Gbps.

\begin{figure}[H]
\centering
\setlength\fheight{5cm}
\setlength\fwidth{7cm}
\definecolor{mycolor1}{rgb}{0.00000,0.44700,0.74100}%
\definecolor{mycolor2}{rgb}{0.85000,0.33000,0.10000}%
\begin{tikzpicture}

\begin{axis}[%
width=0.951\fwidth,
height=\fheight,
at={(0\fwidth,0\fheight)},
scale only axis,
xmin=-10,
xmax=30,
axis x line*=bottom,
axis y line*=left,
xlabel style={font=\color{white!15!black}},
xlabel={\sffamily{Average Transmit Power (dB)}},
ymin=0,
ymax=4,
ylabel style={font=\color{white!15!black}},
ylabel={\sffamily{Capacity (Gbps)}},
axis background/.style={fill=white},
legend style={at={(0.03,0.97)}, anchor=north west, legend cell align=left, align=left, draw=white!15!black,draw=none}
]
\addplot [color=black, line width=1pt]
  table[row sep=crcr]{%
-10	0.00534645086402242\\
-9	0.00672419759559147\\
-8	0.00845486316864161\\
-7	0.0106276321685984\\
-6	0.01335354427455\\
-5	0.0167704546528268\\
-4	0.0210489034834205\\
-3	0.0263989824412511\\
-2	0.033078115699297\\
-1	0.041399665205362\\
0	0.0517420519811371\\
1	0.0645578208874048\\
2	0.0803818188593625\\
3	0.0998372581949646\\
4	0.123638124456062\\
5	0.152586099207021\\
6	0.187560126222657\\
7	0.22949712151597\\
8	0.279362966174881\\
9	0.338114257822409\\
10	0.40665278363576\\
11	0.48577630407695\\
12	0.576130510877066\\
13	0.678167627857865\\
14	0.792116673441054\\
15	0.917969118988978\\
16	1.05548156219815\\
17	1.20419473051662\\
18	1.3634660034522\\
19	1.53251117037792\\
20	1.71045051547497\\
21	1.89635450122632\\
22	2.0892852123653\\
23	2.28833091423115\\
24	2.49263241104222\\
25	2.70140103760657\\
26	2.91392897964606\\
27	3.12959318024197\\
28	3.34785428715355\\
29	3.56825212239262\\
30	3.79039896626935\\
};
\addlegendentry{\sffamily{Exact}}

\addplot [color=black, dotted , line width=1pt]
  table[row sep=crcr]{%
-10	0.00536683826464955\\
-9	0.00675644907235663\\
-8	0.00850586543068288\\
-7	0.0107082501399882\\
-6	0.0134808882170796\\
-5	0.0169714327500381\\
-4	0.0213657679635787\\
-3	0.026897908231847\\
-2	0.0338624601971797\\
-1	0.042630311648098\\
0	0.0536683826464955\\
1	0.0675644907235663\\
2	0.0850586543068288\\
3	0.107082501399882\\
4	0.134808882170796\\
5	0.169714327500381\\
6	0.213657679635787\\
7	0.26897908231847\\
8	0.338624601971797\\
9	0.42630311648098\\
10	0.536683826464955\\
11	0.675644907235663\\
12	0.850586543068288\\
13	1.07082501399882\\
14	1.34808882170796\\
15	1.69714327500381\\
16	2.13657679635787\\
17	2.6897908231847\\
18	3.38624601971797\\
19	4.2630311648098\\
20	5.36683826464955\\
21	6.75644907235663\\
22	8.50586543068289\\
23	10.7082501399882\\
24	13.4808882170796\\
25	16.9714327500381\\
26	21.3657679635787\\
27	26.897908231847\\
28	33.8624601971797\\
29	42.630311648098\\
30	53.6683826464955\\
};
\addlegendentry{\sffamily{Low Power Regime}}

\addplot [color=mycolor2, dash pattern={on 7pt off 2pt on 1pt off 3pt} , line width=1pt]
  table[row sep=crcr]{%
-10	-5.47671060047293\\
-9	-5.26186402552412\\
-8	-5.05298656876538\\
-7	-4.8438433988805\\
-6	-4.61539391152843\\
-5	-4.3948250375883\\
-4	-4.15774665542384\\
-3	-3.92267973556007\\
-2	-3.69179769802253\\
-1	-3.46057051096844\\
0	-3.22726076905803\\
1	-2.99735451269409\\
2	-2.76575046487501\\
3	-2.53341980014264\\
4	-2.30251795237268\\
5	-2.07091253961837\\
6	-1.83930174944687\\
7	-1.60769082674047\\
8	-1.37608006337325\\
9	-1.14446910720901\\
10	-0.912858245917626\\
11	-0.681247386567723\\
12	-0.449636526763503\\
13	-0.218025666458005\\
14	0.0135851936435865\\
15	0.245196053615989\\
16	0.476806913658975\\
17	0.708417774245852\\
18	0.940028633822843\\
19	1.17163949393156\\
20	1.40325035405083\\
21	1.63486121417674\\
22	1.86647207434324\\
23	2.09808293446289\\
24	2.32969379461687\\
25	2.56130465477168\\
26	2.79291551487059\\
27	3.02452637539093\\
28	3.25613723513137\\
29	3.48774809526343\\
30	3.71935895539695\\
};
\addlegendentry{\sffamily{High Power Regime}}

\addplot [color=mycolor1, dash pattern={on 7pt off 2pt on 0pt off 0pt}, line width=1pt]
  table[row sep=crcr]{%
-10	0.00535248518771738\\
-9	0.0067337219412389\\
-8	0.00846988708381465\\
-7	0.0106513113177739\\
-6	0.0133908112907062\\
-5	0.0168289981869839\\
-4	0.0211406745328724\\
-3	0.0265424469445855\\
-2	0.0333016380420479\\
-1	0.0417464906004477\\
0	0.0522774989881148\\
1	0.0653794502836759\\
2	0.0816333830033012\\
3	0.101727157310993\\
4	0.126462686914446\\
5	0.156757175888519\\
6	0.193635091303034\\
7	0.238207361427307\\
8	0.291634802931367\\
9	0.355074444182467\\
10	0.429610423782818\\
11	0.516175218962522\\
12	0.615471119444287\\
13	0.727904549949359\\
14	0.853545510831649\\
15	0.992120410231476\\
16	1.14303980223997\\
17	1.30545525912409\\
18	1.4783342947336\\
19	1.66054044114473\\
20	1.8509072039786\\
21	2.048298393361\\
22	2.25165159551556\\
23	2.46000506244582\\
24	2.67251050217756\\
25	2.88843518780473\\
26	3.10715681728157\\
27	3.32815405334978\\
28	3.55099497905439\\
29	3.7753250228445\\
30	4.00085533724029\\
};
\addlegendentry{\sffamily{Jensen's Upper Bound}}

\end{axis}
\end{tikzpicture}%
    \caption{Capacity versus the average transmit power. Illustration of the exact closed-form, Jensen's bound as well as the lower and upper bounds.}
     \label{fig8}
\end{figure}
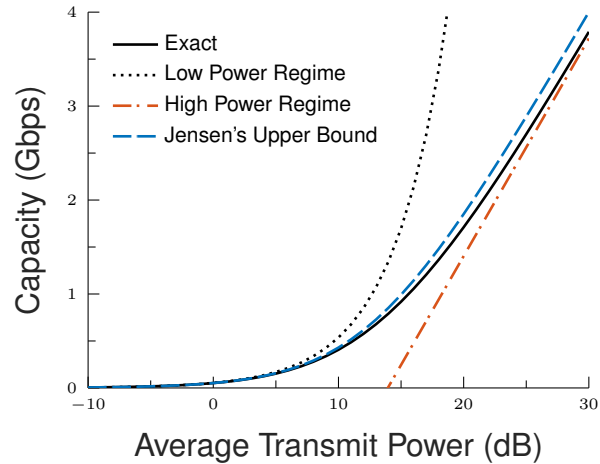

Fig.~\ref{fig8} illustrates the characteristics of the capacity in terms of exact, low and high power regimes, and upper bound expressions. For low power, we observe that the characteristic (green) is perfectly tight to exact expression, however, it substantially deviates up to 15 dB. In addition, the high power characteristic perfectly aligned to the exact expression around 25 dB. Moreover, the gap between the exact and the upper bound is negligible even at high power. Thereby, the three derived characteristics show a good agreement and accuracy relative to the exact closed-form expression.

\begin{figure}[H]
\centering
\setlength\fheight{5cm}
\setlength\fwidth{7cm}
\input{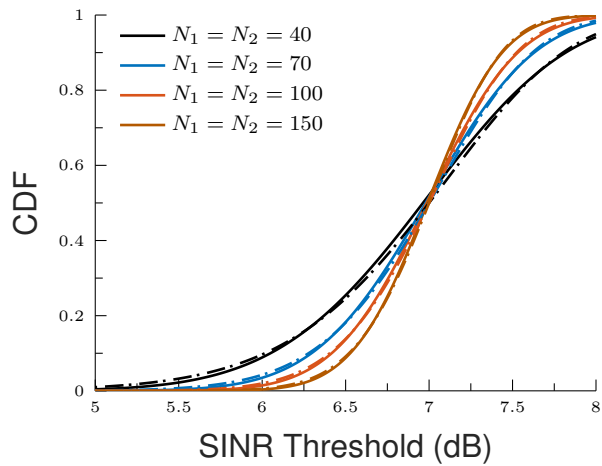}
    \caption{CDF of the overall SINR for different number of antennas. The dashdotted and solid lines correspond to the Gaussian approximation and the exact expressions, respectively.}
     \label{fig9}
\end{figure}
Fig.~\ref{fig9} provides the dependence of the exact and approximate CDF with respect to the SINR threshold for massive numbers of TX and RX antennas. Clearly, as the number of antennas increases, the outage improves. We observe that the Gaussian approximation becomes more accurate relative to the exact expression when the number of antennas is very large. Particularly, the gap between the approximation and the exact expressions for 150 antennas is very small compared to the gap for 40 antennas.
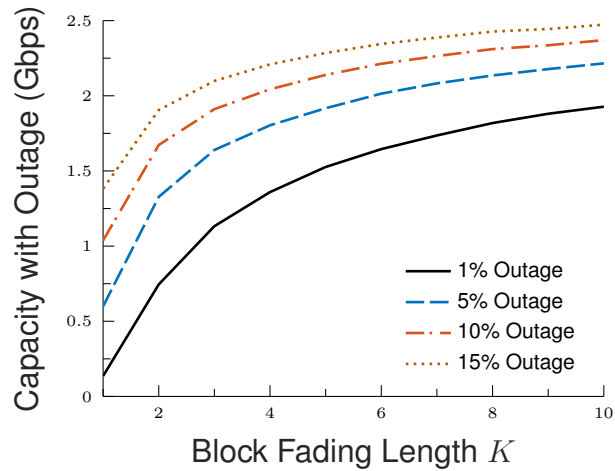
\begin{figure}[H]
\centering
\setlength\fheight{5cm}
\setlength\fwidth{7cm}
\definecolor{mycolor1}{rgb}{0.00000,0.44700,0.74100}%
\definecolor{mycolor2}{rgb}{0.85000,0.33000,0.10000}%
\begin{tikzpicture}

\begin{axis}[%
width=0.951\fwidth,
height=\fheight,
at={(0\fwidth,0\fheight)},
scale only axis,
xmin=1,
xmax=10,
axis x line*=bottom,
axis y line*=left,
xlabel style={font=\color{white!15!black}},
xlabel={\sffamily{Block Fading Length $K$}},
ymin=0,
ymax=2.5,
ylabel style={font=\color{white!15!black}},
ylabel={\sffamily{Capacity with Outage (Gbps)}},
axis background/.style={fill=white},
legend style={at={(0.97,0.03)}, anchor=south east, legend cell align=left, align=left, draw=white!15!black,draw=none}
]
\addplot [color=black, line width=1pt]
  table[row sep=crcr]{%
1	0.135812735688296\\
2	0.745021954437557\\
3	1.13251194495231\\
4	1.35924194652362\\
5	1.52637210722403\\
6	1.64522976669881\\
7	1.73606297391341\\
8	1.81860766388632\\
9	1.88042024827249\\
10	1.92748561599238\\
};
\addlegendentry{\sffamily{1\% Outage}}

\addplot [color=mycolor1, dash pattern={on 7pt off 2pt on 0pt off 0pt}, line width=1pt]
  table[row sep=crcr]{%
1	0.599430509845224\\
2	1.32758672867999\\
3	1.6393845724452\\
4	1.80362615588483\\
5	1.91714929061792\\
6	2.0142660358703\\
7	2.08247824821213\\
8	2.13508413522247\\
9	2.1775307258111\\
10	2.21559153185424\\
};
\addlegendentry{\sffamily{5\% Outage}}

\addplot [color=mycolor2, dash pattern={on 7pt off 2pt on 1pt off 3pt} , line width=1pt]
  table[row sep=crcr]{%
1	1.03760510689574\\
2	1.67239287611032\\
3	1.91116200685283\\
4	2.04310817577201\\
5	2.13891878507073\\
6	2.21296082156771\\
7	2.26429684376411\\
8	2.31055771595756\\
9	2.33593040955899\\
10	2.37001293171144\\
};
\addlegendentry{\sffamily{10\% Outage}}

\addplot [color=orange!70!black, dotted, line width=1pt]
  table[row sep=crcr]{%
1	1.38227846163049\\
2	1.90520950936484\\
3	2.09857388920074\\
4	2.20906713753788\\
5	2.28463471788871\\
6	2.34474579858546\\
7	2.38749922418629\\
8	2.42722741689639\\
9	2.44416703883268\\
10	2.47313866370376\\
};
\addlegendentry{\sffamily{15\% Outage}}

\end{axis}
\end{tikzpicture}%
    \caption{Capacity with outage versus the number of block fading for different outage rates.}
     \label{fig10}
\end{figure}

\begin{figure}[H]
\centering
\setlength\fheight{5cm}
\setlength\fwidth{7cm}
\definecolor{mycolor1}{rgb}{0.00000,0.44700,0.74100}%
\definecolor{mycolor2}{rgb}{0.85000,0.33000,0.10000}%
\begin{tikzpicture}

\begin{axis}[%
width=0.951\fwidth,
height=\fheight,
at={(0\fwidth,0\fheight)},
scale only axis,
xmode=log,
xmin=1e-4,
xmax=1,
xminorticks=true,
axis x line*=bottom,
axis y line*=left,
xlabel style={font=\color{white!15!black}},
xlabel={\sffamily{Outage Probability}},
ymin=0,
ymax=3.5,
ylabel style={font=\color{white!15!black}},
ylabel={\sffamily{Capacity with Outage (Gbps)}},
axis background/.style={fill=white},
legend style={at={(0.03,1)}, anchor=north west, legend cell align=left, align=left, draw=white!15!black,draw=none,fill=none}
]
\addplot [color=black, line width=1pt]
  table[row sep=crcr]{%
0.003653	0.137503523749935\\
0.006876	0.171069138514398\\
0.012559	0.212244743260395\\
0.02334	0.262464707140878\\
0.04098	0.323299322693842\\
0.071587	0.396409161163114\\
0.119072	0.48347495334568\\
0.191701	0.586103926445348\\
0.294603	0.705719050773513\\
0.425374	0.843443825203652\\
0.577175	1\\
0.72739	1.17563663469239\\
0.852586	1.37010466975099\\
0.937191	1.58268235491156\\
0.980168	1.81224619130063\\
0.99577	2.0573732086068\\
0.999443	2.31645617962626\\
0.999961	2.58781437356203\\
0.999998	2.86978721917029\\
1	3.16080442391302\\
1	3.4594316186373\\
};
\addlegendentry{\sffamily{$P_1 = P_2$ = 0 dB}}

\addplot [color=mycolor1, dash pattern={on 7pt off 2pt on 0pt off 0pt} , line width=1pt]
  table[row sep=crcr]{%
0.00013	0.137503523749935\\
0.000245	0.171069138514398\\
0.000523	0.212244743260395\\
0.000959	0.262464707140878\\
0.001868	0.323299322693842\\
0.003602	0.396409161163114\\
0.00675	0.48347495334568\\
0.012607	0.586103926445348\\
0.023033	0.705719050773513\\
0.040693	0.843443825203652\\
0.071225	1\\
0.119445	1.17563663469239\\
0.191521	1.37010466975099\\
0.294603	1.58268235491156\\
0.426371	1.81224619130063\\
0.576749	2.0573732086068\\
0.727121	2.31645617962626\\
0.853514	2.58781437356203\\
0.937186	2.86978721917029\\
0.980298	3.16080442391302\\
0.995843	3.4594316186373\\
};
\addlegendentry{\sffamily{$P_1 = P_2$ = 5 dB}}

\addplot [color=mycolor2, dash pattern={on 7pt off 2pt on 1pt off 3pt} , line width=1pt]
  table[row sep=crcr]{%
5e-06	0.137503523749935\\
1.4e-05	0.171069138514398\\
1.5e-05	0.212244743260395\\
2.7e-05	0.262464707140878\\
8.3e-05	0.323299322693842\\
0.000127	0.396409161163114\\
0.000253	0.48347495334568\\
0.000508	0.586103926445348\\
0.000994	0.705719050773513\\
0.001903	0.843443825203652\\
0.003548	1\\
0.006888	1.17563663469239\\
0.012666	1.37010466975099\\
0.02278	1.58268235491156\\
0.041108	1.81224619130063\\
0.071245	2.0573732086068\\
0.118992	2.31645617962626\\
0.192994	2.58781437356203\\
0.294424	2.86978721917029\\
0.425811	3.16080442391302\\
0.575967	3.4594316186373\\
};
\addlegendentry{\sffamily{$P_1 = P_2$ = 10 dB}}

\addplot [color=orange!70!black, dotted, line width=1pt]
  table[row sep=crcr]{%
1e-06	0.137503523749935\\
0	0.171069138514398\\
0	0.212244743260395\\
0	0.262464707140878\\
1e-06	0.323299322693842\\
3e-06	0.396409161163114\\
6e-06	0.48347495334568\\
2.4e-05	0.586103926445348\\
3.3e-05	0.705719050773513\\
5.2e-05	0.843443825203652\\
0.000148	1\\
0.000264	1.17563663469239\\
0.000532	1.37010466975099\\
0.000968	1.58268235491156\\
0.001921	1.81224619130063\\
0.003604	2.0573732086068\\
0.006786	2.31645617962626\\
0.012714	2.58781437356203\\
0.023146	2.86978721917029\\
0.040295	3.16080442391302\\
0.071063	3.4594316186373\\
};
\addlegendentry{\sffamily{$P_1 = P_2$ = 15 dB}}

\end{axis}
\end{tikzpicture}%
    \caption{Capacity with outage dependence on the outage probability for different transmit powers at the source and the relay.}
     \label{fig11}
\end{figure}
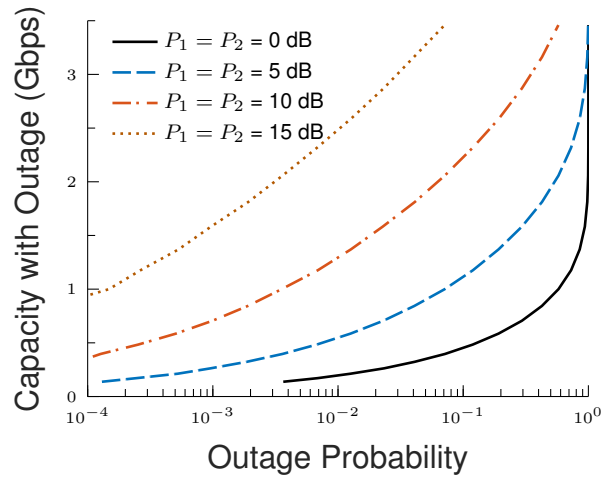

Fig.~\ref{fig10} shows that the capacity improves when the outage rate increases. In fact, the receiver requires a minimum receive SINR to decode the received data. This conditioned receive SINR must guarantee that the system is not in outage. To decode the data, higher outage percentile means large receive SINR to get rid of the outage state. Thereby, the capacity in turn improves because of the high SINR threshold conditioned for successful decoding. In addition, for a small number of block fading, the capacity averaging over the number of block fading is not accurate. However, up to 6 blocks fading, this number is sufficiently enough to get a constant capacity over time.

Fig.~\ref{fig11} illustrates the capacity dependence on the outage probability. We observe that for low transmit power (0 dB), the outage is severe between 1e-2 and 1e-3 yielding lower capacity. However, as the transmit power increases the outage performance improves around 1e-4 resulting in higher capacity. In addition, a higher outage probability results in a good capacity with outage which is already discussed and illustrated by Fig.~\ref{fig10}. This result is also considered and confirmed in \cite[Chapter 4]{gold} and the relative threshold design is discussed in \cite[Chapter 14]{gold}.

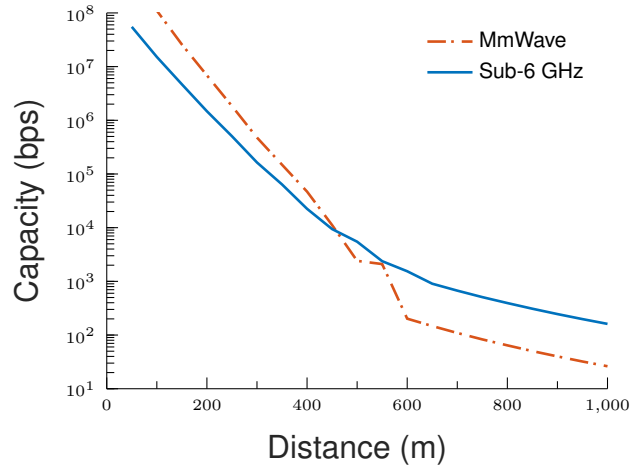
\begin{figure}[H]
\centering
\setlength\fheight{5cm}
\setlength\fwidth{7cm}
\definecolor{mycolor1}{rgb}{0.00000,0.44700,0.74100}%
\definecolor{mycolor2}{rgb}{0.85000,0.33000,0.10000}%
\begin{tikzpicture}

\begin{axis}[%
width=0.951\fwidth,
height=\fheight,
at={(0\fwidth,0\fheight)},
scale only axis,
xmin=0,
xmax=1000,
axis x line*=bottom,
axis y line*=left,
xlabel style={font=\color{white!15!black}},
xlabel={\sffamily{Distance (m)}},
ymode=log,
ymin=10,
ymax=1e8,
yminorticks=true,
ylabel style={font=\color{white!15!black}},
ylabel={\sffamily{Capacity (bps)}},
axis background/.style={fill=white},
legend style={legend cell align=left, align=left, fill=none, draw=white!15!black,draw=none}
]
\addplot [color=mycolor2, dash pattern={on 7pt off 2pt on 1pt off 3pt} , line width=1pt]
  table[row sep=crcr]{%
50	533158210.627184\\
100	110366277.246078\\
150	25850532.5107096\\
200	6874935.67543606\\
250	1844095.10955564\\
300	476874.586923035\\
350	148882.19332766\\
400	46956.2326228307\\
450	11187.1301722002\\
500	2409.2885670043\\
550	2100.06483622702\\
600	201.815080822955\\
650	147.160062908959\\
700	109.369485657779\\
750	82.9701078342816\\
800	64.2563583925281\\
850	50.2299241333178\\
900	40.0228905122814\\
950	32.1510844008461\\
1000	26.2268441194434\\
};
\addlegendentry{\sffamily{MmWave}}

\addplot [color=mycolor1, line width=1pt]
  table[row sep=crcr]{%
50	55039363.6151944\\
100	15150433.8211444\\
150	4686470.3180821\\
200	1474782.32275158\\
250	505390.574842457\\
300	163963.463004446\\
350	63911.0208219712\\
400	22458.3903324867\\
450	9423.04957831455\\
500	5479.13624477236\\
550	2375.84953464462\\
600	1542.67188932873\\
650	904.244486127958\\
700	671.94161120598\\
750	508.77453268941\\
800	394.477851453973\\
850	309.273806553065\\
900	245.851553176384\\
950	198.130177008259\\
1000	161.385991385399\\
};
\addlegendentry{\sffamily{Sub-6 GHz}}

\end{axis}
\end{tikzpicture}%
    \caption{Capacity for mmWaves (28 GHz) and sub-6 GHz (5.8 GHz) for different distances between the TX, RX and the relay.}
    \label{fig12}
\end{figure}
Fig.~\ref{fig12} provides the variations of the capacity for mmWaves and sub-6 GHz with respect to the distances between TX, RX and the relay. For lower distance , the probability of LOS is higher, which means that most of the power are dominated in LOS direction rather than in the scattering NLOS direction. Given that mmWaves exhibits higher bandwidth compared to sub-6 GHz and the LOS component dominates over NLOS, mmWaves achieves higher rate compared to sub-6 GHz. Inversely, for larger distance, the probability of LOS decreases and most of the power is scattered as there is no dominated component or also NLOS components dominate over LOS component. Given that mmWaves has poor penetration than sub-6 GHz, the received power for mmWaves is much smaller compared to sub-6 GHz. Thereby, sub-6 GHz achieves higher rate than mmWaves for larger distance.

\section{Conclusion}
In this work, we proposed a global framework analysis of a mmWaves vehicular relay channel. We investigated the dependence of the reliability metrics on different parameters such as the number of antennas, the time correlation, the modulation, the power interferers, and the blockage density. We showed that the system becomes more resilient to fading by increasing the number of antennas and having a better channel estimation. Also, the performance improves when the receiver becomes less vulnerable to the interferers powers. Moreover, the capacity depends to a large extent on the blockage density, but, it is still in the order of Gbps even for severe blockage. As future directions, we intend to extend this work by considering wideband channel model and full-duplex relaying. Although such relaying scheme improves the capacity compared to half-duplex mode, it is susceptible to the destructive self-interference relaying which degrades substantially the system performance. A good perspective to address this challenging is to design robust precoders and combiners to cancel the self-interference and enhance the system performance reliability.
\appendices
\section{Proof of the $n$-th moment ($\ref{eq15}$)}
The $n$-th moment is defined as 
\begin{equation}{\label{eq47}}
\mathbb{E}[\gamma^n] = \int\limits_{0}^{\infty}\gamma^n f_{\gamma_{\text{eff},i}}(\gamma)d\gamma.   
\end{equation}
Expanding the expression of the PDF ($\ref{eq13}$) and after some mathematical manipulations, the $n$-th moment can be take the following form
\begin{equation}{\label{eq48}}
\mathbb{E}[\gamma^n] = \sum_{p=0}^{N_im_i}\Theta(N_i,m_i,M_{r,i},m_{r,i},p,\rho_i,\overline{\gamma}_i,\overline{\gamma}_{r,i})\int\limits_0^{\infty}x^{a-1}e^{-bx}(1+cx)^{-(M_{r,i}m_{r,i}+p)}dx  
\end{equation}
Where $\Theta(\cdot,\ldots)$ is a multivariate function depending on the parameters between the parentheses. We introduce this function to simplify the derivation and deal with short mathematical expression.

Using the identity \cite[Eq.~(07.35.26.0003.01)]{18}, ($\ref{eq49}$) can be reformulated as follows
\begin{equation}{\label{eq49}}
\mathbb{E}[\gamma^n] = \sum_{p=0}^{N_im_i}\frac{\Theta(N_i,m_i,M_{r,i},m_{r,i},p,\rho_i,\overline{\gamma}_i,\overline{\gamma}_{r,i})}{\Gamma(M_{r,i}m_{r,i})}\int\limits_0^{\infty}x^{a-1}e^{-bx}G_{1,1}^{1,1} \Bigg(cx ~\bigg|~\begin{matrix} 1-M_{r,i}-p \\ 0 \end{matrix} \Bigg)dx. 
\end{equation}
Using the identity \cite[Eq.~(2.24.3.1)]{prud} and after some mathematical manipulations, the $n$-th moment is derived as ($\ref{eq15}$).
\section{Proof of the probability of error per hop ($\ref{eq21}$)}
The probability of symbol error per hop is defined as 
\begin{equation}\label{eq50}
P_{e,i} = \frac{\alpha\sqrt{\beta}}{2\sqrt{2\pi}}\int\limits_0^{\infty}\frac{e^{-\frac{\beta\gamma}{2}}}{\sqrt{\gamma}}F_{\gamma_{\text{eff},i}}(\gamma)d\gamma.
\end{equation}
After expanding the expression of the CDF ($\ref{eq12}$), the probability of error can be taking the following form
\begin{equation}\label{eq51}
\begin{split}
P_{e,i} =& \frac{(1-\rho_i)^{N_im_i}\alpha\sqrt{\beta}}{2\sqrt{2\pi}}\int\limits_0^{\infty}\frac{e^{-\beta x}}{\sqrt{x}}dx - \sum_{n=0}^{N_im_i-1}\sum_{p=0}^{n}\Psi(N_i,m_i,M_{r,i},m_{r,i},n,p,\rho_i,\overline{\gamma}_i,\overline{\gamma}_{r,i},\alpha,\beta)\\&\times~\int\limits_0^{\infty}x^ae^{-bx}(1+cx)^{-(M_{r,i}m_{r,i}+p)}dx
\end{split}
\end{equation}
where $\Psi(\cdot,\ldots)$ is a multivariate function introduced for a reason of simplicity.

After applying the identity \cite[Eq.~(07.35.26.0003.01)]{18}, ($\ref{eq51}$) can be reformulated as follows

\begin{equation}\label{eq52}
\begin{split}
P_{e,i} =& \frac{(1-\rho_i)^{N_im_i}\alpha\sqrt{\beta}}{2\sqrt{2\pi}}\int\limits_0^{\infty}\frac{e^{-\beta x}}{\sqrt{x}}dx - \sum_{n=0}^{N_im_i-1}\sum_{p=0}^{n}\frac{\Psi(N_i,m_i,M_{r,i},m_{r,i},n,p,\rho_i,\overline{\gamma}_i,\overline{\gamma}_{r,i},\alpha,\beta)}{\Gamma(M_{r,i}m_{r,i}+p)}\\&\times~\int\limits_0^{\infty}x^ae^{-bx}G_{1,1}^{1,1} \Bigg(cx ~\bigg|~\begin{matrix} 1-M_{r,i}-p \\ 0 \end{matrix} \Bigg)dx. 
\end{split}
\end{equation}
After applying the identities \cite[Eq.~(3.381.4)]{15}, and \cite[Eq.~(2.24.3.1)]{prud}, the first and the second integrals are solved and the probability of error is obtained as ($\ref{eq21}$).

\section{Proof of the capacity per hop ($\ref{eq26}$)}
The average channel capacity per hop can be expressed as
\begin{equation}\label{eq53}
\mathcal{C}_i = B\int\limits_0^{\infty}\log_2(1+x)f_{\gamma_{\text{eff},i}}(x)dx 
\end{equation}
After expanding the PDF expression ($\ref{eq13}$) and applying the identity \cite[Eq.~(07.35.26.0003.01)]{18}, the capacity can be reformulated as follows
\begin{equation}\label{eq54}
\begin{split}
\mathcal{C}_i =&  \sum_{p=0}^{N_im_i}\Phi(B,N_i,m_i,M_{r,i},m_{r,i},p,\rho_i,\overline{\gamma}_i,\overline{\gamma}_{r,i})\int\limits_0^{\infty}x^{a-1}e^{-bx}\log(1+x)\\&\times~G_{1,1}^{1,1} \Bigg(cx ~\bigg|~\begin{matrix} 1-M_{r,i}m_{r,i}-p \\ 0 \end{matrix} \Bigg)dx     
\end{split}
\end{equation}
where $\Phi(\cdot,\ldots)$ is a multivariate function used to reduce the expression of the capacity.

After using identity $G_{p,q}^{m,n}\Bigg(z^C~\bigg|~\begin{matrix} a_1,\ldots,a_p \\ b_1,\ldots,b_q \end{matrix} \Bigg) = \frac{1}{C}~H_{p,q}^{m,n}\Bigg(z~\bigg|~\begin{matrix} (a_1,C^{-1}),\ldots,(a_p,C^{-1}) \\ (b_1,C^{-1}),\ldots,(b_q,C^{-1}) \end{matrix} \Bigg)$ \cite{foxh} to convert the exponential, the Meijer-$G$ and the logarithm functions into the Fox-$H$ function as follows
\begin{equation}\label{eq55}
e^{-bx} = G_{0,1}^{1,0}\Bigg(bx~\bigg|~\begin{matrix} - \\ 0 \end{matrix} \Bigg) = H_{0,1}^{1,0}\Bigg(bx~\bigg|~\begin{matrix} - \\ (0,1) \end{matrix} \Bigg).\end{equation}
\begin{equation}\label{eq56}
\log(1+x) = G_{2,2}^{1,2}\Bigg(x~\bigg|~\begin{matrix} 1,~1 \\ 1,~0 \end{matrix} \Bigg) = H_{2,2}^{1,2}\Bigg(x~\bigg|~\begin{matrix} (1,1),~(1,1) \\ (1,1),~(0,1) \end{matrix} \Bigg).
\end{equation}
\begin{equation}\label{eq57}
G_{1,1}^{1,1}\Bigg(cx~\bigg|~\begin{matrix} 1-M_{r,i}m_{r,i}-p \\ 0 \end{matrix} \Bigg) = H_{1,1}^{1,1}\Bigg(cx~\bigg|~\begin{matrix} (1-M_{r,i}m_{r,p}-p,1) \\ (0,1) \end{matrix} \Bigg).    
\end{equation}
After these transformations, the capacity can be expressed as
\begin{equation}\label{eq58}
\begin{split}
\mathcal{C}_i =& \sum_{p=0}^{N_im_i}\Phi(B,N_i,m_i,M_{r,i},m_{r,i},p,\rho_i,\overline{\gamma}_i,\overline{\gamma}_{r,i})\int\limits_0^{\infty}x^{a-1}
H_{0,1}^{1,0}\Bigg(x~\bigg|~\begin{matrix} - \\ (0,1) \end{matrix} \Bigg)\\&\times~H_{2,2}^{1,2}\Bigg(x~\bigg|~\begin{matrix} (1,1),~(1,1) \\ (1,1),~(0,1) \end{matrix} \Bigg) H_{1,1}^{1,1}\Bigg(cx~\bigg|~\begin{matrix} (1-M_{r,i}m_{r,i}-p,1) \\ (0,1) \end{matrix} \Bigg)dx.  
\end{split}
\end{equation}
After using the identity \cite[Eq.~(2.3)]{16}, the capacity can be derived as ($\ref{eq26}$).
\section{Proof of the probability of error for the relaying system ($\ref{eq40}$)}
For the overall relaying system, the probability of error is expressed as
\begin{equation}\label{eq59}
P_e = \frac{\alpha\sqrt{\beta}}{2\sqrt{2\pi}}\int\limits_0^{\infty}\frac{e^{-\frac{\beta\gamma}{2}}}{\sqrt{\gamma}}F_{\gamma_e}(\gamma)d\gamma.   
\end{equation}
After expanding the expression of the CDF ($\ref{eq38}$), the probability of error can be reformulated as 
\begin{equation}\label{eq60}
\begin{split}
P_e =&~\frac{\alpha\sqrt{\beta}}{2\sqrt{2\pi}}\int\limits_0^{\infty}\frac{e^{-\frac{\beta x}{2}}}{\sqrt{x}}\left(F_{\gamma_{\text{eff},1}}(x) + F_{\gamma_{\text{eff},2}}(x) - F_{\gamma_{\text{eff},1}}(x) F_{\gamma_{\text{eff},2}}(x)  \right)dx,\\=&~P_{e,1} + P_{e,2} - \frac{\alpha\sqrt{\beta}}{2\sqrt{2\pi}} \int\limits_0^{\infty}  F_{\gamma_{\text{eff},1}}(x) F_{\gamma_{\text{eff},2}}(x) dx =P_{e,1} + P_{e,2} - \mathcal{J}.
\end{split}
\end{equation}
To simplify the derivation and using the identity \cite[Eq.~(07.35.26.0003.01)]{18}, we adopt the following formulation of the CDF per hop $F_{\gamma_{\text{eff},i}}(\cdot)$ as follows
\begin{equation}\label{eq61}
\begin{split}
F_{\gamma_{\text{eff},i}}(x) =& (1-\rho_i)^{N_im_i}\left(1 - \sum_{n_i=0}^{N_im_i-1}\sum_{p_i=0}^{n_i}\zeta(N_i,m_i,M_{r,i},m_{r,i},n_i,p_i,\rho_i,\overline{\gamma}_i,\overline{\gamma}_{r,i})x^{a_i}e^{-b_ix}\right.\\&~\times\left. G_{1,1}^{1,1}\Bigg(c_ix~\bigg|~\begin{matrix} 1-M_{r,i}m_{r,i}-p_i \\ 0 \end{matrix} \Bigg) \right)     
\end{split}
\end{equation}
where $\zeta(\cdot,\ldots)$ is a multivariate function introduced to simplify the expression.

Multiplying the two expressions of the two CDFs and after some mathematical simplifications, the term $\mathcal{J}$ can be written as
\begin{equation}\label{eq62}
\begin{split}
\mathcal{J} =& (1-\rho_2)^{N_2m_2}\frac{\alpha\sqrt{\beta}}{2\sqrt{2\pi}}\int\limits_0^{\infty}\frac{e^{-\frac{\beta x}{2}}}{\sqrt{x}}F_{\gamma_{\text{eff},1}}(x)dx - (1-\rho_1)^{N_1m_1}(1-\rho_2)^{N_2m_2}\frac{\alpha\sqrt{\beta}}{2\sqrt{2\pi}}\sum_{n_2=0}^{N_2m_2-1}\sum_{p_2=0}^{n_2}\\&\zeta(N_2,m_2,M_{r,2},m_{r,2},n_2,p_2,\rho_2,\overline{\gamma}_2,\overline{\gamma}_{r,2})\int\limits_0^{\infty}x^{a_2-\frac{1}{2}}e^{-(\frac{\beta}{2}+b_2)x} G_{1,1}^{1,1}\Bigg(c_2x~\bigg|~\begin{matrix} 1-M_{r,2}m_{r,2}-p_2 \\ 0 \end{matrix} \Bigg) dx \\&+~ (1-\rho_1)^{N_1m_1} (1-\rho_2)^{N_2m_2}\frac{\alpha\sqrt{\beta}}{2\sqrt{2\pi}}\sum_{n_1=0}^{N_1m_1-1}\sum_{p_1=0}^{n_1}\sum_{n_2=0}^{N_2m_2-1}\sum_{p_2=0}^{n_2}
\\&\zeta(N_1,m_1,M_{r,1},m_{r,1},n_1,p_1,\rho_1,\overline{\gamma}_1,\overline{\gamma}_{r,1})\zeta(N_2,m_2,M_{r,2},m_{r,2},n_2,p_2,\rho_2,\overline{\gamma}_2,\overline{\gamma}_{r,2})\\&\int\limits_0^{\infty}x^{a_1+a_2-\frac{1}{2}}e^{-(b_1+b_2+\frac{\beta}{2})x} G_{1,1}^{1,1}\Bigg(c_1x~\bigg|~\begin{matrix} 1-M_{r,1}m_{r,1}-p_1 \\ 0 \end{matrix} \Bigg) G_{1,1}^{1,1}\Bigg(c_2x~\bigg|~\begin{matrix} 1-M_{r,2}m_{r,2}-p_2 \\ 0 \end{matrix} \Bigg) dx,\\=& \mathcal{J}_1 - \mathcal{J}_2 + \mathcal{J}_3. 
\end{split}
\end{equation}
The first term $\mathcal{J}_1$ is straightforward and is equal to
\begin{equation}\label{eq63}
\mathcal{J}_1 = (1-\rho_2)^{N_2m_2}P_{e,1}.    
\end{equation}
After applying the identity \cite[Eq.~(3.381.4)]{15}, the second term $\mathcal{J}_2$ can be further reduced to
\begin{equation}\label{eq64}
\begin{split}
\mathcal{J}_2 =& (1-\rho_1)^{N_1m_1}(1-\rho_2)^{N_2m_2}\frac{\alpha\sqrt{\beta}}{2\sqrt{2\pi}}\int\limits_0^{\infty}\frac{e^{-\frac{\beta x}{2}}}{\sqrt{x}}dx - (1-\rho_1)^{N_1m_1}P_{e,2},\\=& (1-\rho_1)^{N_1m_1}(1-\rho_2)^{N_2m_2}\frac{\alpha\Gamma(0.5)}{2\sqrt{\pi}} - (1-\rho_1)^{N_1m_1}P_{e,2}.    
\end{split}
\end{equation}
The first step is to convert the exponential and the two Meijer-$G$ functions of $\mathcal{J}_3$ into Fox-$H$ functions similarly to $(\ref{eq55})$ and $(\ref{eq57})$. Then applying \cite[Eq.~(2.3)]{16} and after some mathematical manipulations, the third term $\mathcal{J}_3$ is derived.

Finally, after plugging the terms $\mathcal{J}_1,~\mathcal{J}_2$, and $\mathcal{J}_3$ into ($\ref{eq60}$), the probability of symbol error of the overall relaying system can be obtained as ($\ref{eq40}$).
%\section{}
%\section*{Acknowledgment}
\ifCLASSOPTIONcaptionsoff
  \newpage
\fi
\bibliographystyle{IEEEtran}
\bibliography{main}
\end{document}